\documentclass[acmsmall,screen,nonacm]{acmart}
\usepackage[edges]{forest}
\usepackage{url}
\usepackage{xspace}
\usepackage{booktabs}
\usepackage{makecell}
\usepackage{hyperref}
\usepackage{graphicx}
\usepackage{colortbl}
\usepackage{multirow}
\usepackage{subfigure}
\usepackage{tablefootnote}
\usepackage{color, xcolor}
\usepackage{amsthm,amsmath,amsfonts}
\usepackage{pifont}
\usepackage{enumitem}

\newcommand{\numpaper}{105\xspace}

\usepackage[most]{tcolorbox}
\newcommand{\summary}[1]{
\begin{center}
\begin{tcolorbox}[
    boxsep=2pt,      %
    left=2pt,        %
    right=2pt,       %
    top=2pt,         %
    bottom=2pt,
    title={Summary of Findings}, breakable]
{#1}
\end{tcolorbox}
\end{center}
}

\usepackage[normalem]{ulem}

\newcommand{\delete}[1]{}

\AtBeginDocument{
  \providecommand\BibTeX{{
    \normalfont B\kern-0.5em{\scshape i\kern-0.25em b}\kern-0.8em\TeX}}}

\setcopyright{acmcopyright}
\copyrightyear{2024}
\acmYear{2024}
\acmDOI{1}

\acmJournal{TOSEM}
\acmVolume{1}
\acmNumber{1}
\acmArticle{1}
\acmMonth{1}

\begin{document}

\title{Large Language Models for Unit Testing: A Systematic Literature Review}

\author{Quanjun Zhang}
\email{quanjunzhang@njust.edu.cn}
\orcid{0000-0002-2495-3805}
\affiliation{
\institution{School of Computer Science and Engineering, Nanjing University of Science and Technology}
\city{Nanjing}\country{China}\postcode{210094}
}

\author{Chunrong Fang}
\email{fangchunrong@nju.edu.cn}
\orcid{0000-0002-9930-7111}

\author{Siqi Gu}
\email{siqi.gu@smail.nju.edu.cn}
\orcid{0000-0001-5514-6734}

\author{Ye Shang}
\email{yeshang@smail.nju.edu.cn}
\orcid{0009-0000-8699-8075}

\affiliation{
\institution{State Key Laboratory for Novel Software Technology, Nanjing University}
\city{Nanjing}\country{China}\postcode{210093}
}

\author{Zhenyu Chen}
\email{zychen@nju.edu.cn}
\orcid{0000-0002-9592-7022}
\affiliation{
\institution{State Key Laboratory for Novel Software Technology, Nanjing University}
\city{Nanjing}\country{China}\postcode{210093}
}

\author{Liang Xiao}
\email{xiaoliang@mail.njust.edu.cn}
\orcid{0000-0003-0178-9384}
\affiliation{
\institution{School of Computer Science and Engineering, Nanjing University of Science and Technology}
\city{Nanjing}\country{China}\postcode{210094}
}

\begin{abstract}
Unit testing is a fundamental practice in modern software engineering, with the aim of ensuring the correctness, maintainability, and reliability of individual software components.
Very recently, with the advances in Large Language Models (LLMs), a rapidly growing body of research has leveraged LLMs to automate various unit testing tasks, demonstrating remarkable performance and significantly reducing manual effort.
However, due to ongoing explorations in the LLM-based unit testing field, it is challenging for researchers to understand existing achievements, open challenges, and future opportunities.
This paper presents the first systematic literature review on the application of LLMs in unit testing until March 2025.
We analyze \numpaper{} relevant papers from the perspectives of both unit testing and LLMs.
We first categorize existing unit testing tasks that benefit from LLMs, e.g., test generation and oracle generation.
We then discuss several critical aspects of integrating LLMs into unit testing research, including model usage, adaptation strategies, and hybrid approaches.
We further summarize key challenges that remain unresolved and outline promising directions to guide future research in this area.
Overall, our paper provides a systematic overview of the research landscape to the unit testing community, helping researchers gain a comprehensive understanding of achievements and promote future research.
Our artifacts are publicly available at the GitHub repository: \url{https://github.com/iSEngLab/AwesomeLLM4UT}.
\end{abstract}



\maketitle

\section{Introduction}
\label{sec:introduction}
Unit Testing (UT) aims to validate the correctness of individual components within a software system, and plays a crucial role in the software development and testing lifecycle~\cite{fraser2012whole,fraser2012mutation}.
Nowadays, as modern software continues to evolve and support various critical industries, unit testing has become a standardized and even mandatory practice, forming the cornerstone of software quality and reliability~\cite{liu2023towards}.
However, conducting unit testing manually is often time-consuming and labor-intensive for developers and testing professionals.
For example, prior work~\cite{daka2014survey} has shown that developers typically spend more than 15\% of their time writing unit tests.
To alleviate this burden, researchers have devoted considerable efforts to automating the unit testing process, such as test case and assertion generation~\cite{fraser2011evosuite,barr2015oracle}.
Thus, unit testing has long been an active research topic, extensively studied over the past decades and continuously attracting interest from both academia and industry~\cite{runeson2006survey,zakaria2009unit,daka2014survey,wang2024software}.

Recently, \textbf{Large Language Models (LLMs)} have been increasingly applied in \textbf{Software Engineering (SE)},  fundamentally reshaping the research paradigm in the field~\cite{hou2024large,zhang2023survey_se,wang2024software}.
Built on the Transformer architecture, these models are typically pre-trained on large-scale unlabeled corpora to acquire generic language knowledge~\cite{vaswani2017attention}.
Benefiting from their advanced model architecture, vast parameters and extensive training datasets, LLMs have demonstrated remarkable progress in various software development and testing tasks, including code generation~\cite{li2023skcoder,wang2023two,wang2023natural} and program repair~\cite{zhang2024appt,zhang2023gamma,zhang2024pre}.
In the domain of unit testing, the community has witnessed an explosion of studies equipped with LLMs~\cite{zhang2024exploring,wang2024hits,shang2024large,rao2023cat,tang2024chatgpt}, demonstrating notable advantages and indicating a promising future for further research.
 
However, due to the inherent complexity of unit testing and the rapid evolution of LLMs, integrating LLMs into unit testing workflows is a considerably complex undertaking, making it difficult for interested researchers to understand existing work.
For example, existing LLM-based unit testing studies span a wide range of research perspectives (e.g., empirical~\cite{yang2024evaluation,tang2024chatgpt}, technical~\cite{wang2024hits,gu2024testart} studies), testing scenarios (e.g., test generation~\cite{schaefer2024empirical,zhang2024llm} and oracle generation~\cite{zhang2024exploring,dinella2022toga}), and model usage paradigms (e.g., fine-tuning~\cite{shang2024large,zhang2025improving} and prompting engineering~\cite{ni2024casmodatest,nashid2023retrieval}).
Despite the burgeoning interest and ongoing explorations in the field, the literature currently lacks a detailed and systematic review of the applications of LLMs in unit testing.
This gap makes it challenging for researchers to understand the relationship between LLMs and their use in unit testing, and conduct follow-up research.

\textbf{This Work}.
To bridge this gap, we present the first systematic literature review on the deployment of rapidly emerging LLM-based unit testing studies.
Based on this, the community can gain a comprehensive understanding of existing LLM-based unit testing techniques, offering insights into their strengths, weaknesses, and research trends. 
We collect \numpaper{} relevant papers and conduct a comprehensive analysis from both unit testing and LLMs perspectives.
From our analysis, we reveal crucial challenges and outline future research opportunities in the field.
Overall, our work serves as a valuable resource for researchers and practitioners interested in navigating and advancing this rapidly developing area.

\textbf{{Contributions.}}
To sum up, this work makes the following contributions:
\begin{itemize}
    \item \textit{Survey Methodology.}
    We conduct the first systematic literature review of \numpaper{} high-quality APR papers from 2020 to 2025 that utilize recent LLMs to tackle unit testing challenges.
    
    \item \textit{Trend Analysis.}
    We perform a detailed analysis of selected studies in terms of publication trends and distribution of publication venues.

    \item \textit{UT Perspective.}
    We conduct a comprehensive analysis from the perspective of unit testing to understand the distribution of unit testing tasks with LLMs and provide an in-depth discussion about how these tasks are solved with LLMs.
    
    \item \textit{LLMs Perspective.}
    We conduct a comprehensive analysis from the perspective of LLMs to uncover the commonly-used LLMs, the types of prompt engineering, the input of the LLMs, as well as the accompanying techniques with these LLMs.

    \item \textit{{Challenges and Opportunities.}}
    We highlight some crucial challenges of applying LLMs in the unit testing field and pinpoint promising directions for future research.
\end{itemize}  

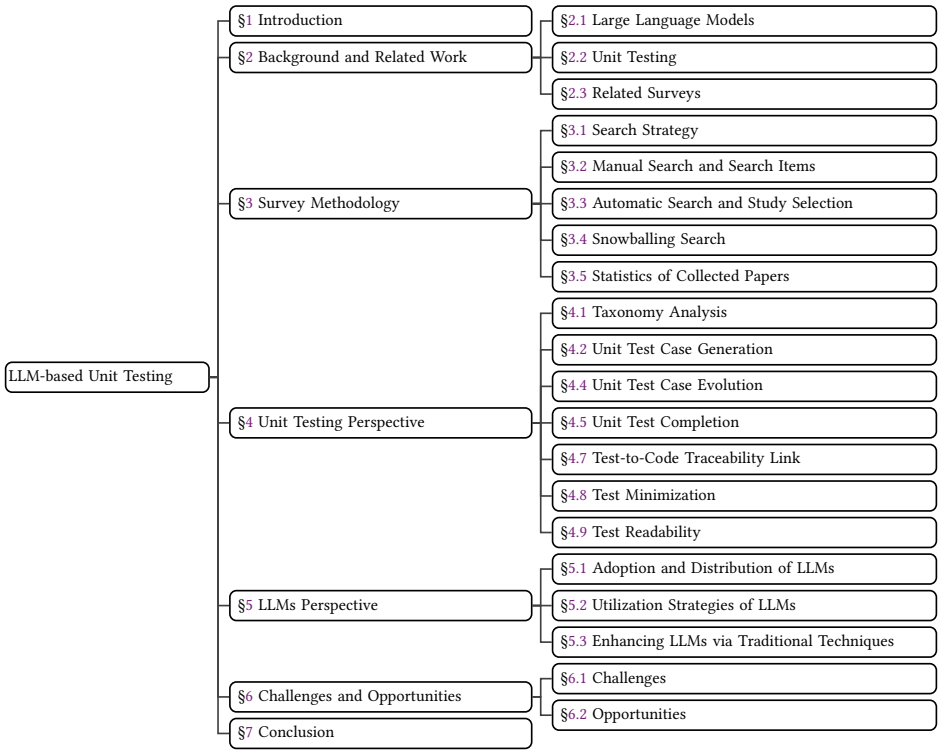
\begin{figure*}[t]
    \centering
    \resizebox{0.9\textwidth}{!}{
        \begin{forest}
            forked edges,
            for tree={
                grow=east,
                reversed=true,
                anchor=base west,
                parent anchor=east,
                child anchor=west,
                font=\large,
                rectangle,
                draw=black,
                rounded corners,
                base=left, 
                align=left,
                minimum width=4em,
                edge+={darkgray, line width=1pt},
                s sep=3pt,
                inner xsep=2pt,
                inner ysep=3pt,
                line width=1pt,
                ver/.style={rotate=90, child anchor=north, parent anchor=south, anchor=center},
            },
            where level=0{text width=12em, font=\normalsize}{},
            where level=1{text width=18em, font=\normalsize}{},
            where level=2{text width=23em, font=\normalsize}{},
            [LLM-based Unit Testing
                [{\;}\S\ref{sec:introduction} Introduction]
                [{\;}\S\ref{sec:background} Background and Related Work
                    [{\;}\S\ref{subsec:llms} Large Language Models]
                    [{\;}\S\ref{subsec:ut} Unit Testing]
                    [{\;}\S\ref{subsec:related_work} Related Surveys]
                ]
                [{\;}\S\ref{sec:methodology} Survey Methodology
                    [{\;}\S\ref{subsec:search_strategy} Search Strategy]
                    [{\;}\S\ref{subsec:search_items} Manual Search and Search Items]
                    [{\;}\S\ref{subsec:study_colection} Automatic Search and Study Selection]
                    [{\;}\S\ref{subsec:snowballing} Snowballing Search]
                    [{\;}\S\ref{subsec:statistics_papers} Statistics of Collected Papers]
                ]
                [{\;}\S\ref{sec:ut_perspective} Unit Testing Perspective
                    [{\;}\S\ref{subsec:ut_analysis} Taxonomy Analysis]
                    [{\;}\S\ref{subsec:test_generation} Unit Test Case Generation]
                    [{\;}\S\ref{subsec:test_evolution} Unit Test Case Evolution]
                    [{\;}\S\ref{subsec:test_completion} Unit Test Completion]
                    [{\;}\S\ref{subsec:test_link} Test-to-Code Traceability Link]
                    [{\;}\S\ref{subsec:test_minimization} Test Minimization]
                    [{\;}\S\ref{subsec:test_readability} Test Readability]
                ]
                [{\;}\S\ref{sec:llm_perspective} LLMs Perspective
                    [{\;}\S\ref{subsec:distribution_llms} Adoption and Distribution of LLMs]
                    [{\;}\S\ref{subsec:utilization_llms} Utilization Strategies of LLMs]
                    [{\;}\S\ref{subsec:enhancing_llms} Enhancing LLMs via Traditional Techniques]
                ]
                [{\;}\S\ref{sec:challenges} Challenges and Opportunities
                    [{\;}\S\ref{subsec:challenges} Challenges]
                    [{\;}\S\ref{subsec:opportunities} Opportunities]
                ]
                [{\;}\S\ref{sec:conclusion} Conclusion]
            ]   
        \end{forest}
        }
    \caption{Structure of this paper.}
    \label{fig:paper_structure}
\end{figure*}

\textbf{{Paper Organization.}}
Figure~\ref{fig:paper_structure} summarizes the structure of this survey.
The remainder of this paper is organized as follows. 
Section~\ref{sec:background} introduces some basic concepts about unit testing and LLMs.
Section~\ref{sec:methodology} illustrates the survey methodology.
Section~\ref{sec:ut_perspective} and Section~\ref{sec:llm_perspective} conduct the analysis from the perspectives of unit testing and LLMs, respectively.
Section~\ref{sec:challenges} discusses key challenges and research guidelines.
Section~\ref{sec:conclusion} draws the conclusions.

\section{Background}
\label{sec:background}
In this section, we introduce the core concepts relevant to this work, including LLMs, unit testing, and related surveys.

\subsection{Large Language Models}
\label{subsec:llms}

LLMs refer to a class of large-scale Transformer-based models that are pre-trained on massive textual corpora to understand and generate human-like text~\cite{chang2024survey}.
To fully leverage the vast amount of unlabeled data, LLMs are typically trained using self-supervised learning objectives, such as masked language modeling~\cite{feng2020codebert}, masked span prediction~\cite{wang2021codet5}, and causal language modeling~\cite{nijkamp2023codegen}.

LLMs are primarily built on the Transformer~\cite{vaswani2017attention} architecture, which consists of an encoder for input representation and a decoder for output generation.
Based on architecture design, LLMs can be categorized into three types: (1) encoder-only models (e.g., BERT~\cite{devlin2019bert} and CodeBERT~\cite{feng2020codebert}), designed for understanding tasks; (2) encoder-decoder models (e.g., T5~\cite{raffel2020exploring} and CodeT5~\cite{wang2021codet5}), designed for translation tasks; and (3) decoder-only models (e.g., LLaMA~\cite{touvron2023llama} and CodeLLaMA\cite{facebook2023codellama}), designed for generation tasks.
In addition, based on their accessibility, LLMs can be categorized into black-box models (e.g., GPT-4~\cite{openai2023gpt4}), which are proprietary and closed-source, and open-source models (e.g., CodeLlama~\cite{facebook2023codellama}), which provide public access to their architecture and weights.
While commercial LLMs continue to dominate the top of the leaderboards, an increasing number of open-source models, such as CodeLlama~\cite{facebook2023codellama} and DeepSeek-R1~\cite{deepseekai2025r1}, are emerging and demonstrating strong performance across a variety of tasks.
Owing to their advanced training mechanisms, model architectures and extensive training datasets, LLMs demonstrate impressive capabilities across a wide range of SE tasks~\cite{zhang2023survey_se,wang2024software}.
In this work, we systematically investigate how LLMs have been applied to address the long-standing challenges of automating unit testing tasks.

\begin{figure}[t]
    \graphicspath{{figs/}}
    \centering
    \includegraphics[width=0.8\linewidth]{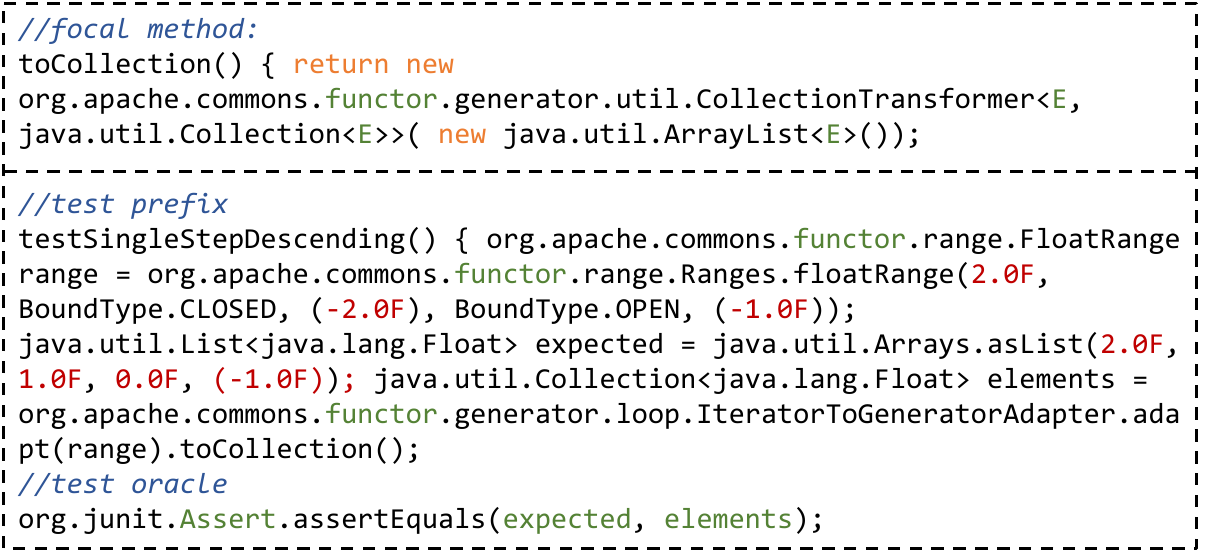}
    \caption{Example of a unit test case}
    \label{fig:unit_example}
\end{figure}

\subsection{Unit Testing}
\label{subsec:ut}
Software testing plays a crucial role in SE by evaluating and ensuring the correctness, reliability, and performance of software systems~\cite{wang2024software}. 
Existing testing practices include unit testing, integration testing, system testing, and acceptance testing, each targeting different stages and levels in the validation process of software systems.
Among them, unit testing is particularly important, as it serves as the foundation for detecting bugs at an early stage of development, thereby facilitating subsequent software development activities~\cite{liu2023towards}.
With its long-established history, unit testing has garnered significant attention from both academia and industry, becoming an accepted and even mandatory practice in modern software engineering.

The primary objective of unit testing is to validate individual components or units of a program by isolating them and executing unit test cases.
This practice helps ensure that each component functions as expected before being integrated with other components of the system.
As illustrated in Figure~\ref{fig:unit_example}, given a focal method (i.e., the unit under test), its unit test typically consists of two components: 
(1) a test prefix, i.e., a sequence of statements that manipulate the unit under test to a specific state, and (2) a test oracle, i.e., an assertion that defines the expected behavior or condition to be satisfied in that state.
In the literature, numerous approaches have been proposed to generate unit test cases automatically, including symbolic execution testing~\cite{cadar2011symbolic,chipounov2011s2e}, random testing~\cite{pacheco2007randoop,ma2015grt}, and search-based testing~\cite {fraser2011evosuite,baresi2010testful}.
In addition, unit testing encompasses a variety of sub-tasks depending on the perspective, such as oracle generation~\cite{zhang2024exploring} and test repair~\cite{yaraghi2024automated}, each addressing specific challenges and scenarios.
These tasks represent distinct aspects of unit testing across various levels of granularity and practical contexts, offering a comprehensive view of unit testing.
Given the complexity and significance of unit testing, this work focuses on how LLMs have been leveraged to support and automate various unit testing tasks.

\subsection{Related Surveys}
\label{subsec:related_work}
\begin{table*}[htbp]
\footnotesize
  \centering
  \caption{Comparison of related surveys and our work}
  \label{tab:related_work}
    \begin{tabular}{l|l|l|l|l|l|l}
    \toprule
    Surveys & Year  & Models & \multicolumn{1}{l|}{Scope} & STR   & Time frame & \# Papers \\
    \midrule
    Watson et al.~\cite{watson2022systematic} & 2022  & DL    & Software Engineering & Y     & 2009-2019 & 128 \\
    Wang et al.~\cite{wang2023machine} & 2022  & ML/DL & Software Engineering & Y     & 2009-2020 & 1428 \\
    Yang et al.~\cite{yang2022survey} & 2022  & DL    & Software Engineering & Y     & 2015-2020 & 142 \\
    Wang et al.~\cite{wang2024software} & 2023  & LLMs  & \multicolumn{1}{l|}{Software Testing} & Y     & 2019-2023 & 102 \\
    Fan et al.~\cite{fan2023large} & 2023  & LLMs  & Software Engineering & \ding{55}     & -     & - \\
    Zheng et al.~\cite{zheng2025towards} & 2023  & LLMs  & Software Engineering & Y     & 2022-2024 & 123 \\
    Hou et al.~\cite{hou2024large} & 2023  & LLMs  & Software Engineering & Y     & 2017-2023 & 395 \\
    Zhang et al.~\cite{zhang2023survey_se} & 2024  & LLMs  & Software Engineering & Y     & 2017-2024 & 1009 \\
    \midrule
    Runeson et al.~\cite{runeson2006survey} & 2006  & -     & Unit Testing & \ding{55}     & -     & - \\
    Zakaria et al.\cite{zakaria2009unit} & 2008  & -     & Unit Testing & \ding{51}     & 2007-2009 & 27 \\
    Dake et al.~\cite{daka2014survey} & 2014  & -     & Unit Testing & \ding{55}     & -     & - \\
    \midrule
    Our Work & 2025  & LLMs  & Unit Testing & \ding{51}     & 2020-2025 & \numpaper{} \\
    \bottomrule
    \end{tabular}%
\end{table*}%

There exist several surveys or literature reviews on the general application of LLMs in the broader area of software engineering~\cite{zhang2023survey_se,zheng2025towards,fan2023large,hou2024large,wang2024software}.
Different from these studies targeting the whole software engineering/testing workflow, this work focuses specifically on the achievements of LLMs in the domain of unit testing, which remains relatively underexplored.
Moreover, unit testing and its related tasks have been widely surveyed in the past~\cite{runeson2006survey,zakaria2009unit,daka2014survey}.
However, these studies primarily focus on traditional unit testing techniques and were conducted prior to 2014, before the emergence of modern LLMs.
Thus, they do not overlap with our research, as the first LLM-based unit testing work emerged in 2020.
Table~\ref{tab:related_work} presents a detailed comparison between our survey and existing literature, highlighting the novelty and scope of our work.
In summary, to the best of our knowledge, this is the first systematic literature review specifically focusing on the applications of LLMs for unit testing.

\section{Survey Methodology}
\label{sec:methodology}
In this section, we describe our methodology for conducting this systematic literature review, including the search strategy, paper collection process, and paper trend analysis.

\subsection{Search Strategy}
\label{subsec:search_strategy}

Following prior SE surveys~\cite{zhang2023survey,wang2023machine}, we adopt a three-stage ``Quasi-Gold Standard'' (QGS) strategy~\cite{zhang2011indentifing} to collect relevant research papers systematically.
This strategy combines manual and automated search processes to construct a set of known relevant studies, which is a common practice to refine search queries and improve retrieval accuracy.

We first conduct a manual search to identify a seed set of relevant papers and derive a search string, as detailed in Section~\ref{subsec:search_items}.
We then utilize the search string to perform an automated search and employ a series of relatively strict filtering steps to extract the most relevant studies, as detailed in Section~\ref{subsec:study_colection}.
We finally use a snowballing search to further complement the search results by manually inspecting references and citations, as detailed in Section~\ref{subsec:snowballing}.
Given the large volume of relevant papers, this strategy allows us to capture the most pertinent papers while maintaining higher efficiency and rigor than a purely manual process.  process.
Particularly, we undertake the following three phases to search for and identify relevant studies.

\subsection{Manual Search and Search Items}
\label{subsec:search_items}

To construct the search items, we conduct a manual search from four top-tier SE conferences (ICSE, ESEC/FSE, ASE, ISSTA) and two journals (TOSEM and TSE), as listed in Table~\ref{tab:se_venues}.
The first two authors independently search these venues and propose an initial set of candidate papers involving both unit testing and LLMs. 
All authors then collaboratively review this collection to finalize the seed set of relevant publications.
Then, we analyze the titles, abstracts, and keywords of these papers to identify search items, and conduct brainstorming to refine our search items, such as synonym substitution.
Finally, this iterative process formulates the set of search items, listed as follows.

\begin{itemize}
    \item \textbf{Search items related to unit testing}:  
    ``unit testing'' OR ``(unit) test cases'' OR ``(unit) test generation'' OR ``(unit) test repair'' OR ``(unit) test evolution'' OR ``regression testing'' OR ``test refactoring'' OR ``test oracle'' OR ``test reduction'' OR ``test selection'' OR ``test readability''

    \item \textbf{Search items related to LLMs}:  
    ``Large Language Model(s)'' OR ``LLM(s)'' OR ``Pre-trained'' OR ``Pretraining'' OR ``PLM(s)'' OR ``(Code)BERT'' OR ``(Code)T5'' OR ``(Code)GPT'' OR ``Codex'' OR ``ChatGPT'' OR ``(Code)Llama'' OR ``GPT-*'' OR ``DeepSeek(*)'' OR ``Mistral''
\end{itemize}

\begin{table*}[t]
  \centering
  \caption{Publication venues for manual search}
  \resizebox{0.98\linewidth}{!}{
    \begin{tabular}{ll}
    \toprule
    \textbf{Acronym} & \textbf{Venues} \\
    \midrule
    ICSE  & International Conference on Software Engineering \\
    ESEC/FSE & Joint European Software Engineering Conference and Symposium on Foundations of Software Engineering \\
    ASE   & International Conference on Automated Software Engineering \\
    ISSTA & International Symposium on Software Testing and Analysis \\
    TOSEM & Transactions on Software Engineering Methodology \\
    TSE   & Transactions on Software Engineering \\
    \bottomrule
    \end{tabular}%
    }
  \label{tab:se_venues}%
\end{table*}%

\subsection{Automatic Search and Study Selection}
\label{subsec:study_colection}

To perform the automated search, we utilize the above search items to collect relevant papers across four widely used databases, i.e., Google Scholar repository, ACM Digital Library, and IEEE Explorer Digital Library, at the end of January 2025.
We restrict our search to articles published from 2017 onward, as the Transformer architecture~\cite{vaswani2017attention}, which serves as the foundation for LLMs, was introduced that year.

\subsubsection{Inclusion and Exclusion Criteria}

We define a set of inclusion and exclusion criteria to filter out papers that do not align with the scope of this survey.

\textbf{Inclusion criteria}.
We define the following criteria for including papers:
\begin{itemize}
    \item \textbf{I1}: The paper proposes a technique, tool, or framework to address unit testing tasks using LLMs.
    \item \textbf{I2}: The paper conducts an empirical study to evaluate LLMs in the context of unit testing.
    \item \textbf{I3}: The paper focuses on specific unit testing tasks (e.g., assertion generation) where LLMs are employed.
\end{itemize}

\textbf{Exclusion criteria}.
We define the following criteria for excluding papers:
\begin{itemize}
    \item \textbf{C1}: The paper does not involve any unit testing tasks, e.g., unit test generation.
    \item \textbf{C2}: The paper does not utilize any LLMs, e.g., only using traditional recurrent neural networks.
    \item \textbf{C3}: The paper only mentions LLMs in future work rather than integrating LLMs in the approach.
    \item \textbf{C4}: The paper is a previously published conference paper extended to a journal by the same authors.
    \item \textbf{C5}: The paper is not an original research study, such as literature reviews, surveys, tool demonstrations, or editorials.
    \item \textbf{C6}: The paper is a duplicate publication where the preprint and the published version have different titles.
    \item \textbf{C7}: The paper is published in a workshop or a doctoral symposium.
    \item \textbf{C8}: The paper is a grey publication, e.g., a technical report or thesis. 
    \item \textbf{C9}: The paper is inaccessible in full text or not written in English.
\end{itemize}

During this process, the first two authors carefully review each paper to determine its eligibility based on the inclusion and exclusion criteria. 
In cases where their decisions differ, the paper is referred to the third author for a final decision.  
For example, following exclusion criterion \textbf{C4}, there exists one study that extends a previously published conference paper~\cite{mastropaolo2021studying} to a journal version~\cite{mastropaolo2023using} by the same authors.
In such cases, we retained only the extended journal version.  
Besides, following exclusion criterion \textbf{C6}, we identify four papers whose published versions have different titles compared to their preprints. 
In such cases, to avoid duplication, we retain only the published versions. After this step, we retained a total of 153 papers.

\subsubsection{Quality Assessment}
To further maximize the inclusion of high-quality papers, we design ten quality assessment questions to evaluate the relevance and rigor of included papers.
For each paper, its quality is evaluated by a three-tier scoring system: criteria are rated as ``yes'' (1 point), ''partial'' (0.5 points), or ``no'' (0 points). 
If a paper accumulates a total score below the threshold of 8 points, it will be excluded from further analysis, which reduces the number of
papers to 99.
The designed quality assessment criteria (QAC) are listed as follows.

\begin{itemize}
    \item \textbf{QA1}. Is the paper primarily focused on LLMs, rather than using them only as baselines?
    \item \textbf{QA2}. Is the paper's impact on the unit testing community explicitly stated?
    \item \textbf{QA3}. Are the research goals and key contributions explicitly defined?
    \item \textbf{QA4}. Has the paper been published in a reputable venue?
    \item \textbf{QA5}. Does the paper provide open-source artifacts for reproducibility, such as datasets, code, or benchmarks?
    \item \textbf{QA6}. Is the implementation of the proposed technique described with sufficient clarity and detail?
    \item \textbf{QA7} Are the experimental settings thoroughly explained, such as including hyperparameters and computing environments?
    \item \textbf{QA8} Are the utilized LLMs explicitly described, along with a clear explanation of how they are applied in the study?
    \item \textbf{QA9}. Are the evaluation metrics and results clearly aligned with research goals?
    \item \textbf{QA10}. Are both the contributions and limitations of the paper critically discussed? 
\end{itemize}

\subsection{Snowballing Search}
\label{subsec:snowballing}

To ensure the completeness of our study, we adopt a snowballing search approach~\cite{watson2022systematic} to manually incorporate papers that are previously overlooked yet remain pertinent to our study.
Specifically, we examine all references (i.e., backward snowballing) or citations (i.e., forward snowballing) of collected papers to assess their quality and relevance to our survey.
The manual inspection continues until no new relevant papers are identified, ultimately leading to the inclusion of an additional 6 papers in our survey.
This rigorous procedure helps ensure that our final corpus is comprehensive and provides a solid foundation for the subsequent analysis of LLM-based unit testing techniques. Finally, we obtain 105 papers that are related to our work.

\subsection{Statistics of Collected Papers}
\label{subsec:statistics_papers}

\begin{figure}[t]
\centering
\begin{minipage}{0.5\textwidth}
    \includegraphics[width=\linewidth]{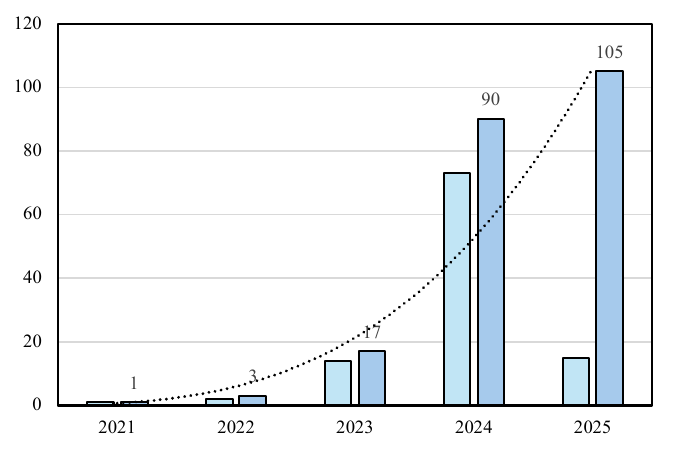}
    \caption{Distribution of papers across years}
    \label{fig:trend_time}
\end{minipage}
\hfill
\begin{minipage}{0.45\textwidth}
    \includegraphics[width=\linewidth]{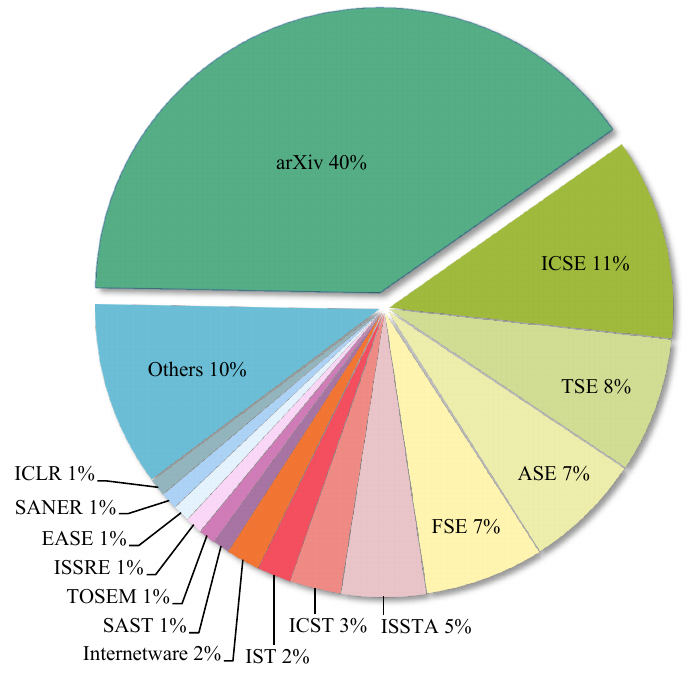}
    \caption{Distribution of papers across venues}
    \label{fig:venues}
\end{minipage}
\end{figure}

Figure~\ref{fig:venues} lists the number of relevant papers published across different publication venues.
We find that 60\% of the papers are published in peer-reviewed venues, with ICSE (11\%) and TSE (8\%) being the most popular conference and journal, respectively.
Following them are ASE (7\%), FSE (7\%), and ISSTA (5\%), all of which are recognized as top-tier software engineering venues.
This trend indicates researchers are increasingly prioritizing the dissemination of their work through high-quality, peer-reviewed venues, which in turn drives innovation and further advances in the field.
Meanwhile, around 40\% of these papers, which are hosted on arXiv, have not undergone peer review.
One possible reason for this phenomenon is the sudden surge of related studies emerging within a brief period.
Given the inconsistent quality of these non-peer-reviewed papers, we carry out a rigorous assessment process to ensure only high-caliber works are selected for this study.

Figure~\ref{fig:trend_time} further illustrates the publication trend of LLM-based unit testing.
We find that the number of relevant papers appears to be rising at an almost exponential pace.
While only one paper was published in 2021 and two in 2022, the number increased to 14 in 2023 and surged to 73 in 2024, with 15 already appearing by March 2025.
Since our data collection ends in March 2025, the figure may not reflect the full trend for the year, and the final count is expected to increase further.
These observations suggest that the application of LLMs to unit testing has gained substantial momentum since 2021 and will likely continue to be an active and growing area of research.

\section{Analysis from Unit Testing Perspective}
\label{sec:ut_perspective}
In this section, we present a comprehensive analysis from the perspective of unit testing and organize the collected studies according to different testing scenarios.

\subsection{Taxonomy Analysis}
\label{subsec:ut_analysis}
\begin{figure}[t]
\centering
    \includegraphics[width=0.7\linewidth]{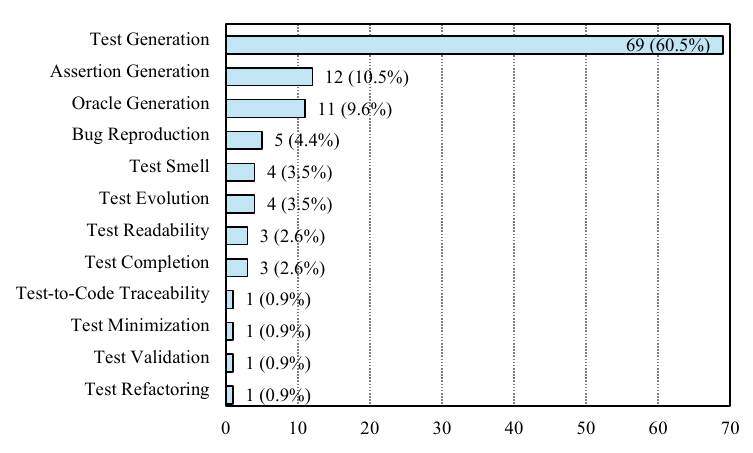}
    \caption{Distribution of unit testing tasks with LLMs}
    \label{fig:ut}
\end{figure}

Figure~\ref{fig:ut} illustrates the distribution of unit testing tasks to which LLMs are applied.
It can be observed that test generation dominates the research landscape, constituting approximately 60\% of the total research volume.
This phenomenon is reasonable, as test case generation has long been a central component in the unit testing pipeline and continues to attract substantial attention from both academia and industry.
In addition, oracle generation represents the most popular task in LLM-based unit testing research, accounting for about 14\% of the research proportion.
This reflects the persistent challenges in automating this task and the growing interest in leveraging LLMs to address it.
Moreover, several other tasks have received moderate attention. 
For example, bug reproduction is explored in 5 studies (4.3\%), while test evolution and test smell detection each appear in 4 studies (3.4\%).  
Tasks such as test completion, test readability, and test minimization are addressed in 1–3 papers each, suggesting they are emerging but less mature application areas for LLMs.
Notably, we observe that some studies investigate underexplored aspects of unit testing, including test-to-code traceability, test refactoring, and test validation, each appearing in only a single study.  
This indicates a growing research interest in broadening the scope of unit testing tasks supported by LLMs thanks to LLMs' general knowledge gleaned from vast amounts of training data.
Overall, this trend highlights both the areas where LLMs have already shown strong potential and the opportunities for further exploration in more specialized or complex unit testing scenarios.

\subsection{Unit Test Case Generation}
\label{subsec:test_generation}

Unit test generation typically takes a focal method (i.e., the method under test) as input and produces a complete unit test comprising two key components: a test prefix and an assertion.
The test prefix sets up the required execution context, while the assertion verifies whether the focal method behaves as expected.
In the literature, to reduce manual efforts in writing unit tests, researchers have proposed various automated unit test generation approaches, including symbolic execution testing~\cite{cadar2011symbolic,chipounov2011s2e}, random testing~\cite{pacheco2007randoop,ma2015grt}, and search-based testing~\cite {fraser2011evosuite,baresi2010testful} strategies.
Traditional approaches (e.g., EvoSuite~\cite{fraser2011evosuite}) are generally designed to maximize code coverage but often struggle to produce meaningful and human-readable test cases, primarily due to their limited understanding of the semantics of focal methods.
Therefore, recent research has turned to leveraging LLMs for test generation with the aim of producing more practical unit test cases from multiple dimensions, including correctness, coverage, and fault detection capability.
Existing LLM-based test generation techniques can be broadly categorized into three distinct groups, discussed as follows.

\subsubsection{Training LLMs for Test Generation}
These techniques typically utilize supervised learning to train LLMs, thereby adapting them to the downstream task of unit test case generation.
This is an intuitive yet effective strategy to enable LLMs to refine their pre-trained knowledge and weight parameters through limited-scale, domain-specific datasets.
In the unit testing domain, the pre-training and fine-tuning paradigm has been extensively adopted during the early stages of LLM development, particularly for medium-scale models such as T5 and CodeT5, which contain hundreds of millions of parameters.
The widespread adoption of this paradigm can be primarily attributed to the limited generalization capabilities of early-stage LLMs, which require task-specific training to achieve optimal performance in specialized domains like test case generation~\cite{tufano2020unit,alagarsamy2024a3test,alagarsamy2024enhancing}.

As early as 2020, Tufano et al.~\cite{tufano2020unit} introduced AthenaTest, the first LLM-based unit test generation approach, which formulates the task as a sequence-to-sequence learning problem.
AthenaTest follows a two-stage training procedure: (1) denoising pre-training on a large, unsupervised Java corpus and (2) supervised fine-tuning for a downstream translation task of
generating unit test cases.
To address the limitation of assertion knowledge in AthenaTest, A3Test~\cite{alagarsamy2024a3test} enhances BART with an assertion-aware self-supervision objective, i.e., predicting the masked tokens of a given focal method and its corresponding assert statements.

Beyond the standard pre-training and fine-tuning pipeline, researchers have explored various strategies to improve the performance of training LLMs in test case generation, including reinforcement learning ~\cite{steenhoek2023reinforcement}, domain adaptation~\cite{shin2024domain} and data augmentation~\cite{he2024data}.
For example, Shin et al.~\cite{shin2024domain} investigate the advantages of domain adaptation for fine-tuning LLMs in the context of automated test case generation. 
However, most of these works rely on general-purpose, off-the-shelf LLMs that are typically trained on natural language and code corpora, without incorporating test-specific knowledge. 
To address this gap, Rao et al.~\cite{rao2023cat} propose CAT-LM, a GPT-style LLM with 2.7 billion parameters specifically trained to learn the mapping between methods under test and their corresponding test cases, making it more suitable for the test generation task.
Similarly, He et al.~\cite{he2024unitsyn} construct a large-scale dataset, UniTSyn, containing 2.7 million method-test function pairs across five programming languages, and propose UniTester, a specialized test generation LLM trained via continual fine-tuning of SantaCoder with an autoregressive objective.

\subsubsection{Prompting LLMs for Test Case Generation}

These techniques typically construct prompts with diverse sources of project information to directly invoke LLMs for test case generation without requiring any additional training.
In the unit testing community, prompt engineering has rapidly gained traction with the emergence of LLMs with billions of parameters, as their advanced capabilities make them well-suited for performing human-like interactions and generating high-quality test cases.

Most prompt-based test generation techniques primarily follow a \textit{generation-and-refinement} paradigm, where initial test cases are first generated based on prompts and then iteratively refined using dynamic execution feedback (e.g., code coverage and failure information) to enhance their quality~\cite{ni2024casmodatest, chen2024chatunitest, gu2024testart, wang2024hits, schaefer2024empirical, yuan2024evaluating, pizzorno2024coverup, ryan2024code}. 
For example, TestART~\cite{gu2024testart} queries LLMs to generate an initial set of test cases, then uses a compiler to collect runtime information. 
This information is then fed into a coverage-guided testing framework and a template-based repair strategy to optimize the test cases iteratively.
In addition, to design more effective prompts, researchers adopt a wide range of strategies to incorporate LLMs with valuable contextual information, including mutation testing~\cite{dakhel2024effective}, method slicing~\cite{wang2024hits}, demonstration retrieval~\cite{zhang2024llm}, defect detection~\cite{yin2024what} and program analysis~\cite{yang2024enhancing}.
For example, Dakhel et al.~\cite{dakhel2024effective} utilize mutation testing to augment the original prompt, enabling LLMs to generate tests capable of killing surviving mutants.
Given a focal method, Yang et al.~\cite{yang2024enhancing} retrieve the most similar method within the same project and incorporate it, along with its corresponding test cases, into the prompt to guide LLMs in generating test cases with project-specific exemplars.

However, most of the aforementioned work has primarily focused on standard benchmarks such as Defects4J~\cite{just2014defects4j} and mainstream programming languages such as Java.  
Recently, the research community has shifted its focus toward emerging frontier domains that present unique and diverse technical challenges. 
These studies include repository-level test case generation~\cite{yin2025enhancing, lops2024system}, multi-language test case generation~\cite{pan2024multi}, high-performance computing software~\cite{karanjai2024harnessing}, industrial deployment~\cite{alshahwan2024automated, sapozhnikov2024testspark}, Rust~\cite{cheng2025rug}, programming problems~\cite{alkafaween2024automating}, data-serialization libraries~\cite{zhong2024advancing}, game development~\cite{paduraru2024unit}, vulnerability exploitation~\cite{gao2024unit}, and bug reproduction~\cite{kang2023large}.  
Overall, these advancements demonstrate the growing versatility of prompt-based LLMs in adapting to diverse testing contexts, without the need for extensive retraining.

\subsubsection{Empirical Study}
In addition to the aforementioned technical advancements, researchers have conducted extensive empirical studies to investigate the capabilities of LLMs in generating unit test cases.
These studies systematically explore the actual performance of LLMs across various aspects, including fine-tuning~\cite{shang2024large,storhaug2024parameter}, prompt engineering~\cite{yang2024evaluation,ouedraogo2024llms,li2024large}, integration or comparison with traditional techniques~\cite{tang2024chatgpt,xiao2024optimizing,abdullin2025test,jiang2024towards,bhatia2024unit}, source code characteristics~\cite{huang2024rethinking}, context~\cite{siddiq2024using}, quality assessment~\cite{taherkhani2024valtest}, retrieval-augmented generation~\cite{shin2024retrieval}, ChatGPT-specific evaluations~\cite{guilherme2023initial,yi2023exploring,yuan2024evaluating} and benchmarking~\cite{bhargava2024cpp,zhang2024testbench,jain2024testgeneval}.
For example, Shang et al.~\cite{shang2024large} conduct a large-scale empirical study to explore the potential of fine-tuning LLMs for unit testing, involving three tasks, five benchmarks, eight evaluation metrics, and 37 advanced LLMs across various architectures and sizes.
Tang et al.~\cite{tang2024chatgpt} perform a systematic comparison of unit test cases generated by ChatGPT and EvoSuite based on several critical factors, including correctness, readability, code
coverage, and bug detection capability.
Yang et al.~\cite{yang2024evaluation} empirically investigate the capabilities of five LLMs with various prompting settings.
These empirical studies provide critical insights into the strengths and limitations of LLM-based test generation and inform future technical design and evaluation.

\subsection{Unit Test Oracle Generation}
\label{subsec:oracle_generation}
\begin{table*}[t]
  \centering
  \caption{Collected studies using LLMs for test oracle automation}
  \resizebox{0.98\linewidth}{!}{
    \begin{tabular}{llrl}
    \toprule
    \textbf{Year} & \textbf{Title} & \textbf{Oracle Type} & \textbf{Ref} \\
    \midrule
    2020  & On Learning Meaningful Assert Statements for Unit Test Cases & Assertion Oracle & \cite{watson2020learning} \\
    2022  & TOGA: A Neural Method for Test Oracle Generation & Whole Oracle & \cite{dinella2022toga} \\
    2022  & Generating Accurate Assert Statements for Unit Test Cases using Pretrained Transformers & Assertion Oracle & \cite{tufano2022generating} \\
    2023  & ChatAssert: LLM-based Test Oracle Generation with External Tools Assistance & Whole Oracle & \cite{hayet2025chatassert} \\
    2023  & Retrieval-Based Prompt Selection for Code-Related Few-Shot Learning & Assertion Oracle & \cite{nashid2023retrieval} \\
    2023  & Using Transfer Learning for Code-Related Tasks & Assertion Oracle & \cite{mastropaolo2023using} \\
    2023  & Neural-Based Test Oracle Generation: A Large-scale Evaluation and Lessons Learned & Whole Oracle & \cite{hossain2023neural} \\
    2024  & Exploring Automated Assertion Generation via Large Language Models & Assertion Oracle & \cite{zhang2024exploring} \\
    2024  & AssertionBench: A Benchmark to Evaluate Large-Language Models for Assertion Generation & Assertion Oracle & \cite{pulavarthi2024assertionbench} \\
    2024  & Chat-like Asserts Prediction with the Support of Large Language Model & Assertion Oracle & \cite{wang2024chat} \\
    2024  & Doc2Oracle: Investigating the Impact of Javadoc Comments on Test Oracle Generation & Whole Oracle & \cite{hossain2024doc2oracle} \\
    2024  & Do LLMs generate test oracles that capture the actual or the expected program behaviour? & Whole Oracle & \cite{konstantinou2024do} \\
    2024  & Transducer Tuning: Efficient Model Adaptation for Software Tasks Using Code Property Graphs & Assertion Oracle & \cite{yusuf2024transducer} \\
    2024  & Deep Multiple Assertions Generation & Assertion Oracle & \cite{wang2024deep} \\
    2024  & Assertify: Utilizing Large Language Models to Generate Assertions for Production Code & Assertion Oracle & \cite{torkamani2024assertify} \\
    2025  & A Large-scale Empirical Study on Fine-tuning Large Language Models for Unit Testing & Assertion Oracle & \cite{shang2024large} \\
    2024  & Towards More Realistic Evaluation for Neural Test Oracle Generation & Whole Oracle & \cite{liu2023towards} \\
    2024  & Assessing Evaluation Metrics for Neural Test Oracle Generation & Whole Oracle & \cite{shin2024assessing} \\
    2024  & An Empirical Study on Focal Methods in Deep-Learning-Based Approaches for Assertion Generation & Assertion Oracle & \cite{he2024empirical} \\
    2025  & TOGLL: Correct and Strong Test Oracle Generation with LLMs & Whole Oracle & \cite{hossain2024togll} \\
    2025  & AugmenTest: Enhancing Tests with LLM-Driven Oracles & Whole Oracle & \cite{khandaker2025augmentest} \\
    2025  & DeCon: Detecting Incorrect Assertions via Postconditions Generated by a Large Language Model & Assertion Oracle & \cite{yu2025decon} \\
    2025  & exLong: Generating Exceptional Behavior Tests with Large Language Models & Exceptional Oracle & \cite{zhang2024generating} \\
    2025  & Improving Retrieval-Augmented Deep Assertion Generation via Joint Training & Assertion Oracle & \cite{zhang2025improving} \\
    2025  & Improving Deep Assertion Generation via Fine-Tuning Retrieval-Augmented Pre-trained Language Models & Assertion Oracle & \cite{zhang2025RetriGen} \\
    \bottomrule
    \end{tabular}%
    }
  \label{tab:oracle}%
\end{table*}%

A test oracle formally defines the expected behavior of a software unit under test for a given test prefix, serving as a critical component in unit testing.
Unlike test case generation, which focuses on producing inputs and exploring execution paths, generating reliable test oracles poses a non-trivial technical challenge, as it requires capturing the intended design specification rather than merely reflecting the implemented behavior.  
This process demands a deep understanding of functional requirements, edge cases, and expected outcomes~\cite{hossain2024doc2oracle}.  
With the advent of LLMs, researchers have begun exploring their potential to automate oracle generation through both fine-tuning and prompting strategies.
Table~\ref{tab:oracle} summarizes existing LLM-based test oracle generation studies, which can be categorized into three key directions: whole oracle generation, assertion oracle generation, and exceptional oracle generation.
We discuss these representative studies in detail below.

\subsubsection{Whole Oracle Generation}
In 2022, Dinella et al.~\cite{dinella2022toga} introduce TOGA, a transformer-based approach to infer both exceptional and assertion test oracles based on the context of the focal method.
TOGA is the first LLM-powered test oracle generation study by fine-tuning CodeBERT to (1) determine whether a test prefix raises an exception and (2) rank a set of candidate assertions.
Furthermore, Liu et al.~\cite{liu2023towards} identify three inappropriate settings in TOGA, i.e., generating test prefixes from correct program versions, evaluating with an unrealistic metric, and lacking a straightforward baseline. 
They then re-evaluate TOGA in a more realistic setting by reducing duplicates and noise during evaluation and introducing an additional ranking step to prioritize failed test cases.  
Similarly, Hossain et al.~\cite{hossain2023neural} conduct a series of replication studies to expand the understanding of TOGA's applicability, generalizability, precision, and fault detection capability across 25 real-world Java systems, 223.5K test cases, and 51K injected faults.

Unlike early-stage TOGA, which fine-tunes CodeBERT as a component for oracle classifier and ranker, recent research explores fine-tuning or prompting LLMs as core backbones to generate oracles in an end-to-end manner.
For example, Hayet et al.~\cite{hayet2025chatassert} introduce CHATASSERT, a feedback-driven oracle generation technique that utilizes prompt engineering to iteratively generate and refine oracles by incorporating dynamic and static information.
Khandaker et al.~\cite{khandaker2025augmentest} propose AugmenTest to generate oracles for EvoSuite-generated test prefixes by prompting LLMs to infer the intended behavior of a focal method from documentation and developer comments.

In addition to the above novel techniques, researchers have conducted numerous empirical studies to explore the capabilities of LLMs in generating oracles.
For example, Hossain et al.~\cite{hossain2024togll} present a comprehensive study by fine-tuning seven code LLMs using six distinct prompts on a large dataset consisting of 110 Java projects.
Hossain et al.~\cite{hossain2024doc2oracle} further conduct an empirical study to investigate the impact of Javadoc comments on test oracle generation by fine-tuning ten LLMs with three different prompts.
Konstantinou et al.~\cite{konuk2024evaluation} empirically investigate whether LLMs can identify the actual and expected program behavior.
Besides, Shin et al~\cite{shin2024assessing} undertake an empirical study to reassess the performance of prior oracle generation techniques (e.g., TOGA) and ChatGPT based on both static generation metrics (e.g., BLEU and CodeBLEU) and dynamic test adequacy metrics (e.g., line coverage and mutation score).
Their findings indicate a lack of statistically significant correlation between static and dynamic metrics, suggesting that existing static generation metrics do not reliably capture the quality of the generated oracles, and dynamic test adequacy metrics should serve as the principal evaluation criteria in this field.

\subsubsection{Assertion Oracle Generation}
In addition to the aforementioned whole test oracle generation work, researchers have conducted specialized explorations on assertion generation, including fine-tuning~\cite{mastropaolo2021studying,mastropaolo2023using,tufano2022generating,tufano2022generating,yusuf2024transducer} and prompt engineering~\cite{nashid2023retrieval,wang2024chat}.
As a seminal work in this domain, Tufano et al.~\cite{tufano2022generating} frame assertion generation as a sequence-to-sequence task by fine-tuning LLMs on the ATLAS dataset, where each input pair consists of a test prefix and its focal method (i.e., the method under test), and the output is an assertion.
This work positions assertion generation as a downstream task for evaluating LLMs in the context of software engineering, laying the groundwork for its adoption in subsequent LLM research~\cite{mastropaolo2021studying,mastropaolo2023using,yusuf2024transducer}.
He et al.~\cite{he2024empirical} conduct an empirical study on the impact of the focal method identification strategy in ATLAS and reveal its limitations in assuming the last method call before assertions as the focal method.
They then introduce ATLAS+, a revised dataset where focal methods are identified using various test-to-code traceability techniques, offering a more realistic and practical evaluation framework.
Unlike the above work relying on training, Nashid et al.~\cite{nashid2023retrieval} utilize few-shot learning to query Codex to generate assertion statements given focal methods and test prefixes.
Wang et al.~\cite{wang2024chat} introduce CLAP, which prompts LLMs to generate assertions based on chain-of-thought reasoning and iteratively refines its predictions through interactions with both LLMs and a Python interpreter.

Recently, Zhang et al.~\cite{zhang2024exploring} conduct the first comprehensive study on fine-tuning various LLMs for automated assertion generation across two benchmarks, five LLMs, and two metrics. 
Their findings underscore the potential of LLM-based assertion generation to substantially alleviate the manual workload of unit testing experts in real-world software development, thereby inspiring further research in this domain.  
Building on this, Zhang et al.~\cite{zhang2025RetriGen} introduce RetriGen, a retrieval-augmented automated assertion generation approach, which incorporates a novel hybrid assertion retriever to refine the assertion retrieval process by leveraging both lexical and semantic similarity to identify the most relevant assertions from external codebases.  
Moreover, Zhang et al.~\cite{zhang2025improving} propose AG-RAG, which optimizes both the retriever and generator within an end-to-end pipeline using a joint training strategy, enabling them to enhance their performance through collaborative learning mutually.

\subsubsection{Exceptional Oracle Generation}
Compared to the extensive research on assertion oracle generation, there has been only one study dedicated specifically to exception oracle generation.
Zhang et al.~\cite{zhang2024generating} introduce exLong, an LLM-based framework for automatically generating exceptional behavior test cases, i.e., checking whether the method under test throws an exception or not.
Built on CodeLlama, exLong integrates program analysis to extract execution traces leading to throw statements, guard conditions, and relevant non-exceptional test cases.

\subsection{Unit Test Case Evolution}
\label{subsec:test_evolution}
During software evolution, source code needs to be continuously changed to satisfy new requirements or fix reported bugs.
To maintain software quality, it is essential to co-evolve the corresponding unit test cases alongside the changed source code.
However, this co-evolution process can pose significant challenges for developers, particularly given the constraints of limited regression testing resources and frequent releases of updated project versions.
To address this, Hu et al.~\cite{hu2023identify} propose CEPROT, which fine-tunes CodeT5 to update outdated test cases based on source code modifications.
Given a source code change and an associated test case, CEPROT first determines whether the test case is obsolete and requires updating; if deemed obsolete, it generates a new version of the test case accordingly.
Building on CEPROT, Liu et al.~\cite{liu2024fix} develop SYNTER, a prompt-based approach that queries GPT-4 using contextual information to generate updated test cases.
Given an obsolete test case to repair, SYNTER constructs three types of test-repair-oriented contexts via static analysis, and ranks them to select the most relevant one for guiding the repair process.
In addition, Yaraghi et al.~\cite{yaraghi2024automated} introduce TARGET, a fine-tuning-based approach that adapts CodeT5+ to more realistic and executable test repair scenarios. 
TARGET frames test repair as a language translation task via a two-step pipeline: collecting essential contextual information that characterizes the test breakage, and utilizing the information to construct input-output training pairs for finetuning LLMs.
Overall, these studies highlight the emerging role of LLMs in automating test evolution, paving the way for more robust and adaptive regression testing in evolving software systems.

\subsection{Unit Test Completion}
\label{subsec:test_completion}
Test completion attempts to automatically generate the next statement within an incomplete unit test case.
This task takes as input the focal method under test, the test method signature, and the preceding statements within the incomplete test body, and produces the subsequent statement to be appended.
TECO~\cite{nie2023learning} represents the first LLM-based test completion approach that leverages static analysis to extract code semantics and fine-tune CodeT5 to predict subsequent test statements.
Furthermore, CAT-LM~\cite{rao2023cat} advances the field by pre-training on both source code and test cases, followed by supervised fine-tuning, demonstrating superior performance over TECO.
This line of work suggests that LLMs can effectively assist in interactive test development workflows by incrementally completing unit test cases.

\subsection{Test Smell}
\label{subsec:test_smell}
Test smells refer to potential issues in unit test cases that may degrade test quality, such as brittleness, slow execution, or lack of clarity.  
Similar to code smells in production code, test smells suggest that a unit test may not be well-structured, robust, or efficient.
Motivated by recent advances in LLMs, Lucas et al.~\cite{lucas2024evaluating} conduct an empirical study to assess the capability of LLMs in detecting test smells, demonstrating their potential to automate this process and improve testing efficiency.
Furthermore, Gao et al.~\cite{gao2024context} propose UTRefactor, an LLM-based test refactoring approach to eliminate test smells and improve the quality of unit test cases.
UTRefactor extracts relevant contextual information from the test code, incorporates external knowledge, and guides LLMs through a chain-of-thought process to simulate manual refactoring with improved accuracy and consistency.
These studies suggest that LLMs can serve as valuable assistants in detecting and mitigating test smells, contributing to more maintainable and robust unit test cases.

\subsection{Test-to-Code Traceability Link}
\label{subsec:test_link}
Test-to-code traceability involves establishing explicit links between unit tests and the corresponding software units they are intended to validate.  
This practice is essential in unit testing, as it provides visibility into how tests align with production code, thereby facilitating test coverage analysis, debugging, and test maintenance.
To this end, Sun et al.~\cite{sun2024method} introduce TestLinker, a hybrid approach that combines heuristic rules with LLMs to establish test-to-code traceability links at the method level. 
Specifically, TestLinker fine-tunes CodeT5 to learn the inherent semantic correlation between unit tests and focal methods, and then utilizes mapping rules to accurately to accurately align predicted function names with corresponding production method declarations.
Despite its importance, this area remains under-explored, calling for further research on LLM-based techniques for test-to-code traceability to enable more maintainable and traceable unit test cases.

\subsection{Test Minimization}
\label{subsec:test_minimization}

As software evolves, unit test cases tend to grow when software evolves, making it impractical to execute all test cases with the allocated testing budgets, especially for large software systems.
Test minimization attempts to improve the efficiency of unit testing by removing redundant test cases, thus reducing execution time and resource consumption while maintaining the adequacy criteria of the test suite, such as code coverage and fault detection capability.
To this end, Pan et al.~\cite{pan2024ltm} propose LTM, a scalable and black-box test minimization approach based on LLMs and similarity analysis.
LTM utilizes LLMs to extract test case embeddings and two measures to compute their similarity via two metrics, i.e., Cosine Similarity and Euclidean Distance.
Based on the similarity, LTM applies a genetic algorithm to optimize the test minimization search space, which identifies the most effective subset of the original test cases within a given testing budget.
However, research on LLM-based test minimization remains limited, suggesting the need for further exploration in broader regression testing scenarios such as test case prioritization and selection.

\subsection{Test Readability}
\label{subsec:test_readability}
Traditional automated unit test generation techniques, particularly search-based tools (e.g., EvoSuite), are capable of producing test cases with high code coverage. 
While these tools alleviate the burden of writing unit tests manually, the generated test cases often suffer from poor readability, making them difficult for developers to understand, interpret, or maintain.
Thus, improving the readability of automatically generated test cases has therefore emerged as an important research direction.
To address this, Delijouyi et al.~\cite{deljouyi2024leveraging} introduce UTGen, which integrates LLMs into the search-based test generation process, thus combining the strengths of both paradigms to generate effective and understandable test cases.
UTGen utilizes LLMs to provide contextually relevant test data, insert informative comments, and suggest suitable variable names via prompt engineering.
Similarly, Biagiola et al.~\cite{biagiola2024improving} employ LLMs to improve the readability of test cases produced by EvoSuite, specifically focusing on renaming variables and test methods, while preserving functional correctness and coverage.
In addition, Zhou et al.~\cite{zhou2024llm} introduce C3, a readability measurement tool that leverages LLMs to extract context-aware readability requirements from source code, aiming to assess and improve the readability of test inputs, especially for primitive and string types.  
Specifically, C3 captures the expected input context from the tested code and checks the consistency of test inputs, and its integration into EvoSuite enables the generation of more readable test inputs by guiding the test generation with these extracted contexts.
These studies demonstrate the potential of LLMs in bridging the gap between traditional unit test generation techniques and human-oriented test comprehension.

\summary{
From the perspective of unit testing, our systematic analysis reveals the following key findings:  
(1) LLMs have been applied across a wide range of unit testing tasks, demonstrating strong potential in both fundamental and emerging scenarios;  
(2) despite this breadth, research efforts remain heavily concentrated on a few primary tasks, with test case generation and oracle generation being the most studied, accounting for 20\% and 19\% of the collected papers, respectively;  
(3) although emerging tasks such as test evolution, test smell detection, and test readability are gaining attention due to LLMs' general-purpose reasoning capabilities, they remain relatively underexplored.
}

\section{Analysis from LLM Perspective}
\label{sec:llm_perspective}

In this section, we perform an analysis of collected papers from the viewpoint of LLMs, specifically focusing on the distribution of LLMs used, their utilization strategies, and traditional techniques employed in conjunction with LLMs.

\begin{figure*}[t]
\centering
    \includegraphics[width=0.9\linewidth]{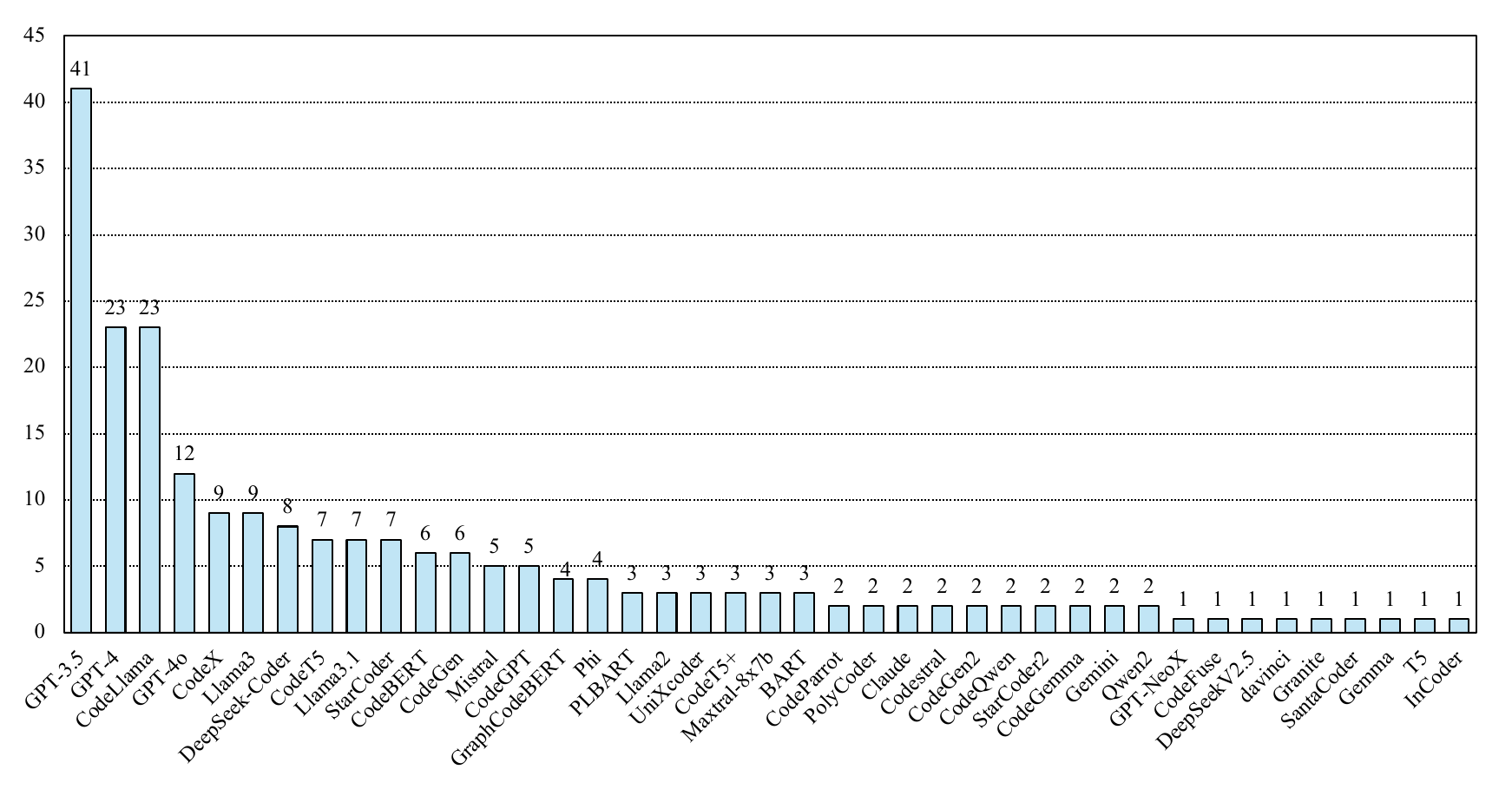}
    \caption{Distribution of LLMs utilized in collected papers}
    \label{fig:llms}
\end{figure*}

\subsection{Adoption and Distribution of LLMs}
\label{subsec:distribution_llms}
Figure~\ref{fig:llms} presents the distribution of LLMs adopted across the collected studies.
The results reveal a clear preference for a few dominant models, particularly those accessible via commercial platforms, while open-source alternatives are gaining traction for their flexibility and adaptability.

Among all models, GPT-3.5~\cite{gpt3.5}, released by OpenAI in November 2022, is the most commonly used LLM in the context of unit testing.  
It is trained using a combination of supervised learning and reinforcement learning from human feedback, which enables it to generate fluent, human-like responses.
Primarily featured in OpenAI's ChatGPT platform, its release has significantly advanced research on LLMs in unit testing, thus contributing to its top ranking in our collected studies.
Following GPT-3.5, GPT-4~\cite{openai2023gpt4} is the second most commonly used LLM in our collected studies.
Released in March 2023, GPT-4 features a much larger parameter count and superior performance across a wide range of tasks than GPT-3.5.
The third-ranked LLM is CodeLlama~\cite{facebook2023codellama}, which is an open-sourced model developed by Meta.
Due to its open-source nature, researchers can not only perform prompt learning, similar to the GPT series models mentioned above, but also conduct additional training to meet domain-specific requirements.
For example, Shang et al.~\cite{shang2024large} fine-tune CodeLlama to support three key unit testing tasks, i.e., test generation, test evolution, and assertion generation.
In addition, GPT-4o~\cite{gpt4o}, the successor to GPT-4, ranks fourth overall and has already been adopted by 12 studies since its release in May 2024. 
Compared with GPT-4, GPT-4o provides notable improvements in response speed, computational efficiency, and long-context reasoning, motivating some studies to explore the potential of CoT in unit testing.
Other less frequently used LLMs include a range of open-source models such as DeepSeek-Coder~\cite{guo2024deepseekcoder}, CodeT5~\cite{wang2021codet5}, CodeBERT~\cite{feng2020codebert}, and StarCoder~\cite{li2023starcoder}.
These open-source models offer flexibility for customization and experimentation, particularly in scenarios that require fine-tuning or integration with task-specific workflows.

\subsection{Utilization Strategies of LLMs}
\label{subsec:utilization_llms}
LLMs are typically pre-trained on large-scale corpora to acquire general-purpose knowledge. 
Thus, a fundamental research challenge arises when integrating off-the-shelf LLMs with unit testing: \textit{how to effectively adapt general-propose LLMs to the specialized tasks of unit testing}.
In this section, we systematically categorize and analyze existing strategies for adapting LLMs to unit testing tasks.

\subsubsection{Taxonomy Analysis}
\begin{figure*}[t]
\centering
    \includegraphics[width=0.99\linewidth]{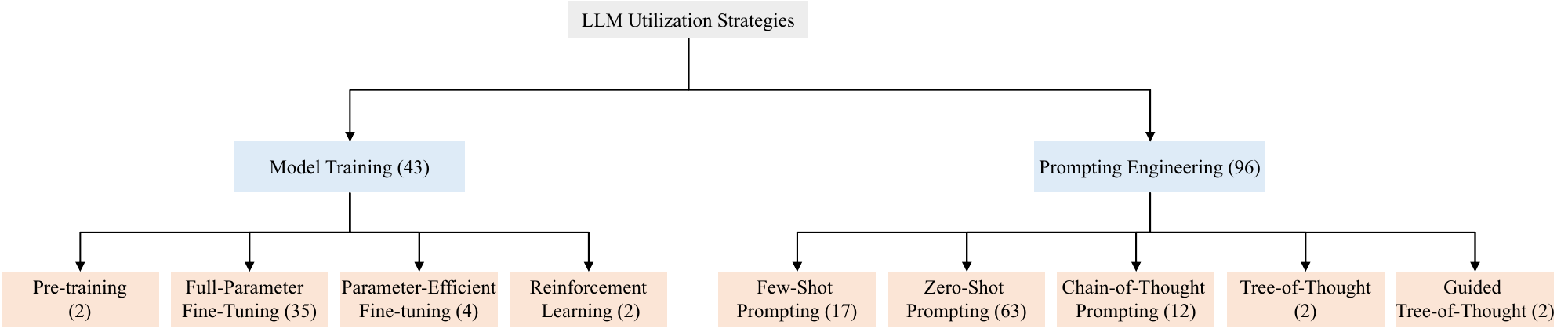}
    \caption{Taxonomy of model utilization strategies in LLM-based unit testing research}
    \label{fig:llm_usage}
\end{figure*}

Figure~\ref{fig:llm_usage} illustrates a hierarchical taxonomy of LLM utilization strategies in unit testing research, annotated with the number of studies adopting each approach.
These adaptation strategies can be broadly grouped into two main categories: \textit{model training} and \textit{prompt engineering}, each comprising several specific techniques that vary in complexity, resource demands, and levels of task alignment.
Overall, our analysis reveals that prompt engineering dominates the landscape with 96 studies, reflecting a growing trend toward leveraging LLMs without additional training.
Zero-shot prompting (63 studies) and few-shot prompting (17 studies) are the most widely adopted strategies, while more advanced prompting techniques such as chain-of-thought (12), tree-of-thought (2), and guided tree-of-thought (2) highlight the growing sophistication of prompt engineering.   
On the training side, full-parameter fine-tuning remains a significant strategy, with 35 studies adapting model weights for unit testing tasks, indicating continued interest in tailoring model weights for specific unit testing objectives.
Although pre-training (2 studies) and reinforcement learning (2 studies) are less commonly used, they demonstrate potential for more targeted or optimization-aware adaptation. 
Parameter-efficient fine-tuning (PEFT), while still emerging (4 studies), represents a practical direction toward efficient adaptation with minimal cost.
In the following, we delve into each strategy in detail, discussing their technical characteristics, representative approaches, and observed trends in LLM-based unit testing research.

\subsubsection{Model Training}
To support task-specific adaptation of LLMs in unit testing, various training strategies have been proposed and evaluated across the literature.  
These strategies differ in terms of training objectives, resource requirements, and expected performance gains.  
We categorize and summarize the main approaches into four types: (1) pre-training models on code-specific data, (2) full-parameter fine-tuning, (3) parameter-efficient fine-tuning (PEFT), and (4) reinforcement learning.  
Each of these strategies represents a distinct point in the trade-off space between model generalization, adaptation cost, and task effectiveness.

\textbf{Model Pre-training}.
In the early stage, inspired by the success of LLMs in natural language, researchers attempt to adapt them to programming languages. 
A straightforward approach is to train code-related LLMs with similar model architectures and training strategies but using code datasets rather than natural language datasets.
For example, to explore the potential of transfer learning in code-related tasks, Mastropaolo et al.~\cite{mastropaolo2021studying} pre-train a T5 model with 1.5M Java methods and fine-tune it to generate assertion statements based on given focal methods and test prefixs.
Furthermore, Rao et al.~\cite{rao2023cat} introduce CAT-LM, a 2.7B-parameter GPT-style LLM pre-trained on aligned focal methods and test cases, making it well-suited for accurate whole-test generation.

\textbf{Full-Parameter Fine-tuning}.
These techniques utilize supervised learning to train LLMs, thereby adapting them to the downstream test case generation task.
This is an intuitive and effective way to allow LLMs to refine their pre-trained knowledge representations and weight parameters through limited-scale, domain-specific datasets.
In the UT community, the training paradigm is extensively adopted during the early stages of LLMs with millions of parameters, such as T5 and CodeT5.
The prevalent implementation of this paradigm can be primarily attributed to the limited generalization capabilities of early-stage LLMs, which require task-specific training to achieve optimal performance in specialized domains like test case generation and assertion generation~\cite{tufano2020unit,alagarsamy2024a3test,alagarsamy2024enhancing}.
We observe that out of 30 studies, LLMs are fine-tuned using full parameter fine-tuning techniques to adapt to downstream unit testing tasks.
For example, AthenaTest~\cite{tufano2020unit} is the first attempt at fine-tuning LLMs for test generation, inspiring a multitude of subsequent studies~\cite{alagarsamy2024a3test,shin2024domain}.
Similar fine-tuning-based studies on other tasks include test evolution~\cite{hu2023identify,yaraghi2024automated}, assertion generation~\cite{he2024empirical}, and test completion~\cite{nie2023learning}.
Besides, Zhang et al.~\cite{zhang2024exploring} empirically explore the potential of fine-tuning LLMs in generating assertions, and Shang et al.~\cite{shang2024large} conduct an empirical study on fine-tuning LLMs to support three unit tasks, i.e., test generation, test evolution, and assertion generation.

\textbf{Parameter-Efficient Fine-tuning (PEFT)}.
However, it is quite computationally expensive and resource-intensive to fine-tune LLMs, particularly for models with billions of parameters, which usually demand substantial GPU resources.
To this end, researchers introduce several parameter-efficient fine-tuning strategies in the domain of unit testing, including low-rank adaptation~\cite{zhang2024generating,bhargava2024cpp,hossain2024doc2oracle}, (IA)$^3$~\cite{storhaug2024parameter} and prompt tuning~\cite{storhaug2024parameter}.
These studies aim to adapt LLMs with minimal weight updates, substantially reducing training costs.
For example, Storhaug et al.~~\cite{storhaug2024parameter} conduct the first empirical study to extensively investigate the effectiveness of PEFT strategies, including LoRA, (IA)$^3$, and prompt tuning, for LLM-based unit test generation.
The results demonstrate that PEFT techniques, particularly LoRA, achieve comparable performance to full-parameter fine-tuning while significantly reducing computational costs, making task-specific adaptation of LLMs more feasible.
Such approaches make LLM-based unit testing more accessible and practical, particularly for academic or industrial settings with limited GPU resources.

\textbf{Reinforcement Learning (RL)}.
In addition to improving fine-tuning efficiency through PEFT, some researchers also employ reinforcement learning to enhance the effectiveness of LLM-based unit testing research~\cite{steenhoek2023reinforcement,takerngsaksiri2024TDD}.
For example, Steenhoek et al.~\cite{steenhoek2023reinforcement} design a lightweight static analysis-based reward mode to analyze the quality of LLM-generated test cases.
They then utilize the reward mode to guide the reinforcement learning process, optimizing LLMs to generate unit tests that maximize expected reward across five code quality metrics.
In addition, Takerngsaksiri et al.~\cite{takerngsaksiri2024TDD} propose PyTester, an RL-based test generation approach for Python in test-driven development settings.
PyTester is optimized with proximal policy optimization with a reward function that incorporates three feedback signals: syntax correctness, test executability, and code coverage.

\subsubsection{Prompt Engineering}
In addition to model training, prompt engineering has become a widely adopted strategy to guide LLM behavior for unit testing tasks.  
These strategies range from simple zero-shot prompts to structured, multi-agent reasoning frameworks.  
We categorize existing prompting strategies in the context of unit testing as follows.

\textbf{Few-shot Prompting}.
Few-shot prompting provides LLMs with several input-output pairs (i.e., demonstrations) before prompting them to generate a response for new, unseen queries. 
For example, in assertion generation, Nashid et al.~\cite{nashid2023retrieval} retrieve similar focal methods along with their corresponding assertions based on embedding-based or frequency-based similarity, using them as in-context examples for Codex.
In addition, for test generation, to provide high-quality and diverse examples, Ni et al.~\cite{ni2024casmodatest} cluster all candidate examples based on semantic similarity and select one representative from each cluster.
The final prompt is then constructed with five examples, which are ordered by three different strategies: random, ascending, and descending order of cosine similarity with the target focal method.

\textbf{Zero-shot Prompting}.
Zero-shot prompting directly queries LLMs to handle unseen tasks without providing any explicit examples.  
This prompt strategy entirely relies on LLMs' pre-existing knowledge and reasoning capabilities, thus requiring a powerful foundation model (such as ChatGPT) for effective interaction.
For example, Gu et al.~\cite{gu2024testart} prompt ChatGPT to generate test cases by defining its role as a unit test case generator.
The prompt also specifies the testing requirements for JDK 1.8 and JUnit 4, as well as a quality requirement to ensure branch coverage in the given focal method.
In oracle generation, Konstantinou et al.~\cite{konstantinou2024do} design a prompt that begins with a role-playing introduction and a task explanation, concluding with a sentence that specifies the desired format of the LLM's response.

\textbf{Chain-of-thought (CoT)}.
CoT prompting attempts to improve LLMs' reasoning by decomposing complex problems into a sequence of intermediate reasoning steps. 
For example, Yang et al.~\cite{yang2024enhancing} utilize CoT reasoning to structure the test generation process into two crucial stages: 
(1) providing LLMs with context information about the focal method to summarize its functionality, thus gaining a deep understanding of the method semantics;
(2) guiding LLMs to iteratively generate divergent test cases that maximize branch coverage by incorporating counter-examples, thus allowing LLMs to reason step by step.
Similarly, Wang et al.~\cite{wang2024hits} instruct LLMs via a step-by-step CoT reasoning process, where LLMs sequentially decompose focal methods, generate test cases, and iteratively refine errors, leading to more effective and higher-coverage unit test generation.
For example, during the method decomposition, LLMs first summarize the focal method's functionality, then recite the meaning and usage of all invoked non-local variables and methods, and finally break down the method into step-based meaningful slices.

\textbf{Tree-of-Thought (ToT)}.  
This is an advanced reasoning technique that enhances the problem-solving capabilities of LLMs by structuring reasoning as a tree-based process.  
Unlike CoT, which follows a linear step-by-step reasoning chain, ToT enables LLMs to iteratively explore and evaluate multiple reasoning paths, discarding suboptimal ones and converging on a more robust solution.  
For example, in the context of unit test generation, Ou{\'{e}}draogo et al.~\cite{draogo2024large} employ ToT by simulating a collaborative process among three virtual software testing experts, where each expert independently proposes test cases, exchanges them with the group for evaluation, and refines them iteratively based on collaborative insights.  
This approach enhances test case quality by leveraging multiple reasoning pathways, ensuring coverage of typical use cases, edge cases, and exception-handling scenarios before merging the finalized test suite.  

\textbf{Guided Tree-of-Thoughts (GToT)}.  
GToT is an enhanced version of ToT that incorporates structured, step-by-step guidance mechanisms to steer the reasoning process toward more systematic and optimal solutions.  
For example, Ou{'{e}}draogo et al.~\cite{draogo2024large} extend ToT by explicitly instructing LLMs to follow a structured framework, including method extraction, functional test generation, edge case identification, and iterative refinement.
In this setting, three virtual experts not only propose and refine test cases but also systematically analyze method signatures and exception-handling scenarios before merging their insights into a complete JUnit 4 test suite.
Besides, by enforcing a well-defined reasoning structure, GToT improves test coverage, enhances code correctness, and minimizes test smells compared to the standard ToT approach.

\begin{figure*}[t]
\centering
    \includegraphics[width=0.7\linewidth]{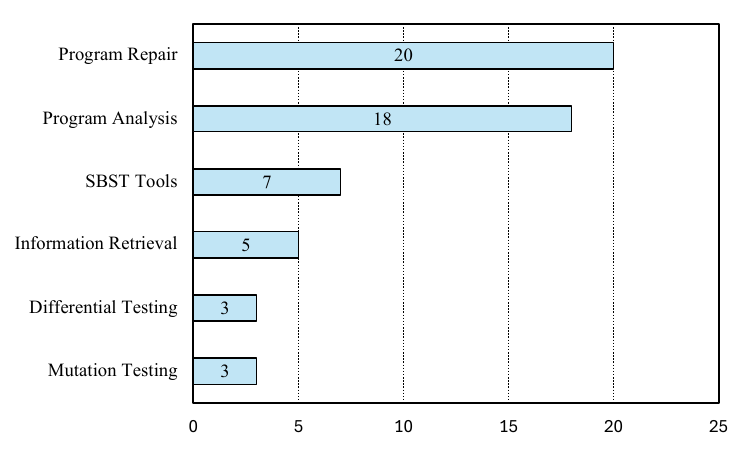}
    \caption{Distribution of integrated traditional techniques in LLM-based unit testing research}
    \label{fig:traditional_techniques}
\end{figure*}

\subsection{Enhancing LLMs via Traditional Techniques}
\label{subsec:enhancing_llms}

Although LLMs have demonstrated remarkable capabilities in various unit testing tasks, the complexity of unit testing poses unique challenges that often exceed the standalone capabilities of LLMs.
To bridge this gap, recent studies have explored hybrid approaches that integrate LLMs with traditional software engineering techniques, aiming to harness the strengths of both paradigms.

\subsubsection{Taxonomy Analysis}
Figure~\ref{fig:traditional_techniques} presents a taxonomy of traditional techniques integrated with LLMs in the context of unit testing, annotated with the number of studies adopting each method. 
These studies fall into six major categories based on the supporting techniques employed: 
\textit{program repair} (20 studies), \textit{program analysis} (18 studies), \textit{SBST tools} (7 studies), \textit{information retrieval} (5 studies), \textit{mutation testing} (3 studies), \textit{differential testing} (3 studies).
Among them, program repair and program analysis are the most frequently adopted techniques, indicating a strong emphasis on improving the correctness and contextual relevance of LLM-generated test cases.
Program repair techniques are primarily used to fix compilation and runtime errors in generated test code, while program analysis enhances prompt relevance by extracting meaningful structural or semantic information. 
Other techniques, such as information retrieval and SBST tools, offer complementary support by providing dynamic context or guiding search-based exploration. 
Mutation and differential testing, though less frequently employed, play important roles in improving the fault-detection capabilities of generated tests.
In the following, we examine each category in detail, discussing its technical motivation, representative approaches, and the observed impact on LLM-based unit testing research.

\textbf{LLMs with Program Analysis}.
To facilitate LLMs' comprehension of the unit under test, an intuitive strategy is to provide them with all relevant information as prompts to the greatest extent possible.
However, this strategy inevitably introduces irrelevant noise in lengthy prompts and hampers LLMs' ability to extract essential semantic information due to the lost-in-the-middle phenomenon.
To this end, researchers utilize program analysis techniques to represent the unit under test and its context information more effectively, thus increasing LLMs' ability to comprehend source code accurately.
For example, in the context of test generation, TELPA~\cite{yang2024enhancing} performs backward and forward program analysis to feed LLMs with a limited number of relevant invocation methods rather than the whole focal class.
Chen et al.~\cite{chen2024chatunitest} conduct program slicing to extract representative API usage and perform dependency analysis to build object construction graphs, which can provide guidance for LLM to generate meaningful test cases.
Wang et al.~\cite{wang2024hits} parse the focal method via static analysis to extract context information for LLMs to understand the usage and structure of the focal method.
In test evolution, Yaraghi et al.~\cite{yaraghi2024automated} utilize static analysis via the Spoon library to construct method-level and class-level call graphs of the focal class, enabling the identification and prioritization of relevant repair contexts for LLMs.
In assertion generation, Wang et al.~\cite{wang2024deep} use Tree-sitter to convert the focal method into a data flow graph, which is then concatenated with the source code sequence as input for GraphCodeBERT.

\textbf{LLMs with Information Retrieval}.
LLMs are typically trained on vast datasets, retaining learned knowledge in the form of parameters up to a fixed cutoff point.
While fine-tuning provides a viable mechanism for incremental knowledge integration, frequently updating LLMs with the latest data is often impractical due to the vast number of model parameters.
As a result, when generating test cases, LLMs may struggle with outdated knowledge and a lack of project-specific context. 
In particular, LLMs may lack awareness of new knowledge (such as updated libraries or frameworks) after their last training date and fail to incorporate critical project-level information (such as method invocation of the unit under test), which reduces the quality of generated test cases and leads to hallucinations.
To address this challenge, researchers leverage information retrieval techniques to dynamically provide LLMs with relevant, up-to-date context information through prompts.
For example, Nashid et al.~\cite{nashid2023retrieval} construct prompts for assertion generation by retrieving demonstrations similar to the test prefix based on embedding or frequency analysis.

\textbf{LLMs with Program Repair}.
Although LLMs have achieved impressive performance on unit testing, generating syntactically correct test code remains challenging, often resulting in compilation and runtime errors.
To address this, researchers have developed various strategies to identify and repair errors in generated test cases automatically.  
Broadly, these program repair techniques fall into two categories: dynamic feedback-based repair~\cite{nie2023learning} and static pattern-based repair~\cite{alagarsamy2024a3test}. 
For example, Yuan et al.~\cite{yuan2024evaluating} adopt a validate-and-repair paradigm, iteratively refining buggy test cases by re-prompting ChatGPT with compilation error messages.
In contrast, inspired by the advancements of traditional templates program repair~\cite{zhang2023gamma}, Gu et al.~\cite{gu2024testart} design five expert-informed repair templates to fix common compilation errors (such as syntax, import, and scope errors) as well as runtime errors in generated test cases.

\textbf{LLMs with Mutation Testing}.
Most LLM-based unit testing approaches have demonstrated promising performance in maximizing code coverage.
However, while code coverage is widely regarded as a useful metric, its correlation with actual bug detection capability remains weak.
Thus, researchers employ mutation testing to simulate real bugs to optimize the bug detection ability of test cases generated by LLMs.
For example, MuTAP~\cite{dakhel2024effective} prompts LLMs to generate initial test cases via zero-shot and few-shot learning.
MuTAP then applies mutation testing to assess how well the generated test cases detect faults (i.e., kill mutants).
If any mutants survive, MuTAP iteratively augments the initial prompt with surviving mutants to re-prompt LLMs to generate improved test cases until all mutants are detected or no further improvements can be made.
Similarly, ACH~\cite{foster2025mutation} attempts to construct mutants that represent faults that are both relevant to the issue at hand and undetected by existing test cases.
These mutants are then used as prompts for LLMs to generate new test cases capable of detecting them.

\textbf{LLMs with Differential Testing}.
Differential testing attempts to detect inconsistencies or bugs by running multiple implementations of a program with the same inputs and comparing their outputs.
Given a program under test, Li et al.~\cite{li2023nuances} prompt ChatGPT to infer the program's intention and ask ChatGPT to generate multiple compilable programs that share the same intention as the original. 
They then apply differential testing between the program under test and the ChatGPT-generated programs, using the inferred intention to identify failure-inducing test cases.
Similarly, Liu et al.~\cite{liu2024llm} first utilize LLMs to generate multiple program variants and test inputs.
They then execute test inputs on both the program under test and its variants to construct the test oracle.
Zhong et al.~\cite{zhong2024advancing} prompt LLMs to generate high-quality unit test cases for JSON library and adopt differential testing to detect potential bugs by comparing the results from fastjson and fastjson2.

\textbf{LLMs with SBST Tools}.
Another line of work combines LLMs with traditional search-based software testing (SBST) tools to leverage their respective strengths.
For example, TELPA~\cite{yang2024enhancing} utilizes the traditional search-based tool Pynguin to generate initial test cases for easily reachable branches, which are then utilized to construct prompts for LLMs, allowing them to address harder-to-cover branches further.
Similarly, when Pynguin fails to increase test coverage within a given testing budget, CODAMOSA~\cite{lemieux2023codamosa} invokes LLMs to generate new seed tests, which are used to resume the generation process of Pynguin.
CoverUp~\cite{pizzorno2024coverup} further extends CodaMosa by incorporating feedback-driven refinements based on coverage information to improve test case quality iteratively.

\summary{
From the perspective of LLMs, our analysis reveals the following key findings:  
(1) the research landscape on LLMs for unit testing is dominated by a few commercial models, with GPT-3.5 and GPT-4 being the most widely utilized, while open-source alternatives such as CodeLlama are gaining attention for flexible adaptation; 
(2) prompt engineering is the most popular adaptation strategy, particularly zero-shot and few-shot prompting, although other advanced reasoning approaches, such as chain-of-thought and tree-of-thought, are emerging but remain relatively underutilized;  
(3) while full-parameter fine-tuning remains popular, there is increasing interest in more cost-effective alternatives, including parameter-efficient fine-tuning (e.g., LoRA) and reinforcement learning;
(4) a growing number of hybrid studies that integrate LLMs with traditional techniques, such as program analysis, mutation testing, and SBST tools, highlight a trend toward improving LLMs with long-standing software engineering practices.
}

\section{Challenges and Opportunities}
\label{sec:challenges}

In this section, we discuss key challenges and highlight potential research directions across both technical and practical dimensions.

\subsection{Challenges}
\label{subsec:challenges}
Despite the promising progress in applying LLMs to unit testing, our survey reveals several persistent challenges that hinder their broader adoption and effectiveness.  
These challenges span from technical limitations in modeling complex software units to broader concerns around dataset quality, bug detection reliability, and model deployability.  
We summarize the key challenges observed in the literature as follows.

\subsubsection{Challenges in Testing Complex Units Under Test}
Unit testing aims to validate the functionality of individual software units in isolation.  
However, in real-world software systems, these units rarely operate independently, and they often rely on complex interactions with other functions, classes, or modules.
This interconnected nature of modern codebases introduces significant challenges for LLM-based unit testing, particularly in understanding and reasoning over the broader execution context.
Our analysis reveals three development stages in how contextual information is incorporated into LLMs for unit testing.
From our collected papers, we observe that early studies typically feed the focal method into LLMs, prompting them to generate corresponding test cases in the form of machine translation.
However, this strategy lacks critical contextual details, such as variable declarations and invoked methods, which are often necessary to produce valid and meaningful test cases.
Later, with the advent of larger models featuring expanded input windows, researchers incorporate the entire focal class as additional context to capture inter-class dependencies.
Although this strategy enriches the available information, it also introduces excessive input size and noise, making it difficult for LLMs to identify and focus on relevant elements. 
Moreover, even this expanded input fails to capture cross-file or cross-module dependencies that are common in large-scale projects.
Recently, some efforts have sought to mitigate these issues by leveraging program analysis techniques to extract semantically related methods, function call relationships, and usage contexts of the unit under test. 
While this strategy improves the relevance of provided information, striking a balance between sufficient context and manageable input size remains an open challenge, especially in repository-level unit testing scenarios.
Future research may explore more advanced static and dynamic analysis techniques (e.g., dataflow analysis) to better identify and structure relevant context. 
Besides, program reduction techniques, such as dead code elimination, can be employed to remove source code that is irrelevant to the behavior of the unit under test, thereby reducing complexity while preserving the essential functionality.

\subsubsection{Challenges in Detecting Real-world Software Bugs}

\begin{figure}[htbp]
\centering
    \includegraphics[width=0.8\linewidth]{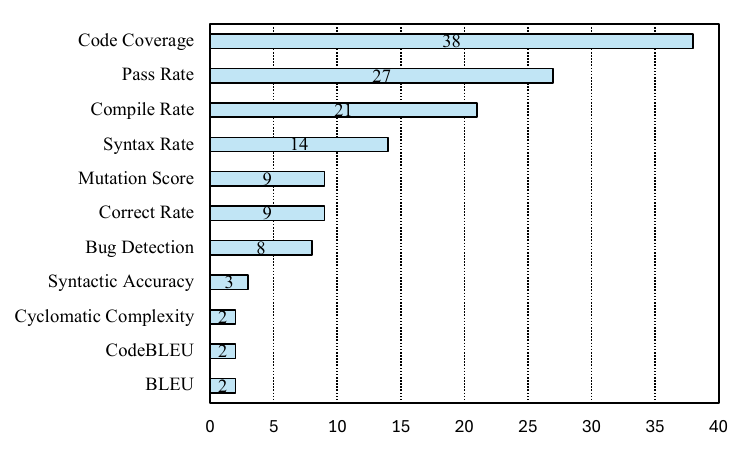}
    \caption{Distribution of evaluation metrics in our collected test generation papers}
    \label{fig:metrics}
\end{figure}

A fundamental purpose of unit testing is to detect individual component bugs at the early stage of software development, thereby improving software reliability.
However, current evaluation metrics for unit testing primarily emphasize generation accuracy, code coverage, and defect detection capabilities, often overlooking its impact on bug detection.
As illustrated in Figure~\ref{fig:metrics}, we summarize the distribution of evaluation metrics adopted in test generation studies based on our collected papers.
It can be observed that the vast majority of studies evaluate the quality of generated test cases using code coverage (38 papers), pass rate (27 papers), and compilation rates (21 papers).
There are only eight papers to identify whether the generated test cases can uncover real-world bugs.

After a careful analysis, we conclude that the challenges of existing unit testing tools in detecting real-world bugs are threefold.
First, the primary reason why current studies lack sufficient focus on bug detection is the limitation of the dataset.
For example, existing studies typically utilize projects from GitHub as evaluation datasets, which usually fail to include real-world bugs.
These datasets guide researchers to evaluate the correctness of test cases (such as syntactic correctness) rather than bug detection capabilities.  
To address this, six papers employ mutation testing to simulate bugs for evaluation.
To address this gap, we call for the development of more comprehensive and realistic datasets tailored for evaluating LLM-driven unit testing approaches.
Second, oracle generation has long remained a persistent challenge in various software testing activities, including unit testing.
For example, Yang et al.~\cite{yang2024evaluation} report that GPT-4 generates valid unit test cases for only 15.74\% of bugs (65/413), and among them, only 40\% (39/65) are correctly detected, resulting in a final detection rate of 9.44\% (39/413).
Similarly, Zhang et al.~\cite{zhang2024exploring} demonstrate that assertions generated by CodeT5 are able to detect only 15 bugs for 835 bugs from Defects4J-v2.0.
The third challenge arises from the complexities inherent in real-world bug detection scenarios.
Although some studies (e.g., Yang et al.~\cite{yang2024evaluation}) attempt to evaluate bug detection, their assessment scenarios lack realism.
For example, they typically generate test cases on the fixed program and run them on the buggy program.
However, in practice, developers expect unit testing tools to uncover bugs in the buggy program and its fixed version is not available at that time.
Besides, test cases generated by LLMs on the fixed program version may implicitly contain some information related to the patch of the bug, which should be unknown during bug finding and results in information leakage issues.
Therefore, to be consistent with the real-world usage
scenario, test cases should be generated on the buggy program
version instead of the bug-fixed program version.
A crucial challenge lies in the issue of false positives, i.e., determining whether a failing test case is caused by its assertion error or by an actual bug in the program.
These false positives hinder the practical application of unit testing techniques in real-world scenarios.

\subsubsection{Challenges in Developing Unit Testing-driven LLMs}
As illustrated in Figure~\ref{fig:llms}, GPT-series models are the most widely adopted in the domain of unit testing. 
However, the proprietary and black-box nature of these commercial models poses several challenges for real-world deployment, including concerns related to deployment environments and potential privacy breaches. 
For example, due to concerns regarding data privacy, when designing LLM-based unit testing tools, most organizations tend to avoid using commercial LLMs in production workflows.
Instead, they often prefer open-source alternatives that can be fine-tuned with domain-specific data in controlled environments.  
However, the computational cost of fine-tuning large-scale models remains prohibitive for most companies, leading them to rely on medium-sized models.  
However, these models often fail to match the performance of their commercial counterparts, as observed in our collected studies.
Among existing LLMs, CAT-LM~\cite{rao2023cat} stands out as the only model specifically designed for test case generation in relation to the method under test. 
However, it still relies on traditional NLP pre-training objectives and has a relatively smaller parameter scale compared to other models.  

These observations point to an urgent need for the development of unit testing-oriented LLMs.  
Therefore, we advocate for the development of more unit test-oriented LLMs. 
However, developing such LLMs presents inherent challenges due to the unique characteristics of unit testing.
First, while software repositories host abundant production code, the number of high-quality, labeled unit test cases remains limited, posing a significant barrier to large-scale pre-training. 
Second, designing pre-training objectives tailored specifically to unit testing, such as coverage maximization and assertion inference, requires further research and innovation to ensure models can generalize effectively in real-world testing scenarios.

\subsection{Opportunities}
\label{subsec:opportunities}

\subsubsection{Opportunities in Developing End-to-End Testing-and-Debugging Framework}
While unit testing primarily aims to detect software bugs at the early stage of development, and program debugging focuses on automatically analyzing and fixing such bugs, the two tasks are inherently connected and mutually reinforcing.  
In particular, high-quality unit test cases not only expose defects, but also support key debugging activities such as root cause analysis, fault localization, patch generation, and patch validation.  
Despite this close relationship, prior work has largely treated unit testing and program debugging as separate research areas, limiting opportunities for integrated solutions and cross-domain enhancement.

In the future, with LLMs serving as the backbone, their powerful understanding and reasoning capabilities of LLMs present an opportunity to bridge this gap, enabling the integration of a fully automated framework for unit testing and program debugging. 
In such a framework, LLMs could first be used to automatically generate test cases and meaningful oracles to uncover potential defects in real-world programs. 
Upon identifying faults, program debugging components could be triggered to perform tasks such as root cause analysis, fault localization, and patch generation.
These patches can then be validated against test cases and reintegrated into the unit testing process, forming a self-reinforcing feedback loop that continuously enhances both testing and debugging effectiveness.
This direction attempts to unify two well-known research domains that are often developed in isolation into an interactive pipeline thereby benefiting the whole software development lifecycle.
More importantly, such integration not only broadens the application scope of these two research areas but also boosts the capabilities of recent LLMs in advancing software quality assurance.

\subsubsection{Opportunities in Exploring More Unit Testing Tasks}
As discussed in Section~\ref{subsec:ut}, a majority of collected papers are concentrated on a limited number of unit testing scenarios, particularly test case generation and oracle generation.
However, several important unit testing scenarios remain largely underexplored in the context of LLMs.  
For example, there is only one paper on the use of LLMs in test minimization, and still no research on test prioritization and test selection.
These tasks typically differ from more widely studied ones, posing unique challenges when integrated with LLMs.
For example, test case prioritization requires analyzing and ranking tens of thousands of test cases based on various features to determine an optimal execution order.  
Unlike the well-explored test generation task, where LLMs address it as a machine translation task by mapping a focal method to its corresponding test code, test prioritization feeds an entire test suite into LLMs, which far exceeds the context window limitations of LLMs.
To address this issue, a promising direction is to combine LLMs with traditional unit testing techniques.  
Specifically, it is promising to leverage LLMs' code semantic understanding capabilities by encoding test cases into vector representations, which are then combined with similarity-based test prioritization algorithms.

\subsubsection{Opportunities in Integrating Multi-modal Context Information}
From our collected papers, in the early stage of LLM-based unit testing research~\cite{tufano2020unit,alagarsamy2024a3test}, LLMs are typically provided with the focal method under test and tasked with directly generating the corresponding test case. 
Later, due to the dependencies among various units, studies attempt to adopt traditional techniques, such as program analysis and information retrieval, to enrich prompts with more relevant contextual information, such as invoked functions and variable definitions.  
However, most existing studies remain largely focused on providing accurate code-level context, often overlooking other valuable input types, such as documentation, API references, and bug reports.
Given the advanced natural language understanding capabilities of LLMs, there is great potential to extract semantic intent and behavioral expectations from these textual sources, which may not be explicitly reflected in the code.  
Moreover, for certain types of software, such as Android applications, the multimodal capabilities of LLMs can be further explored to process diverse inputs, such as graphical user interfaces (GUIs), to support richer and more effective unit testing.  
We believe that the integration of source code and other multi-model contexts (such as textual and visual information) represents a promising direction for improving the completeness and accuracy of LLM-driven unit testing research.

\section{Conclusion}
\label{sec:conclusion}

Unit testing is a crucial and even mandatory practice in modern software engineering, facilitating software development and maintenance.
Recently, Large Language Models (LLMs) have brought transformative capabilities to the unit testing domain, yielding impressive progress and indicating substantial potential in follow-up research.
In this paper, we conduct the first systematic literature review of LLM-based unit testing, covering publications from 2020 to March 2025.  
We analyze \numpaper{} relevant studies from two complementary perspectives: the unit testing dimension, which includes tasks such as test generation, oracle generation, and test evolution; and the LLM dimension, which examines model usage, adaptation strategies, and integration with traditional techniques.  
We also reveal that despite promising progress, challenges remain in areas such as testing complex units, detecting real-world bugs, and developing test-oriented LLMs.
To guide future research, we highlight several research opportunities, including building end-to-end testing-and-repair pipelines, expanding support for underexplored tasks, and leveraging multimodal context information.
Overall, this survey will serve as a comprehensive reference for researchers and practitioners, and help advance the development of LLM-based unit testing research.
\section*{Acknowledgments}
This work is supported partially by the National Natural Science Foundation of China (61932012, 62141215, 62372228) and Frontier Technologies R\&D Program of Jiangsu (BF2024070).

\bibliographystyle{ACM-Reference-Format}
\bibliography{reference}

\end{document}